\documentclass{ws-ijmpa}
\usepackage[utf8]{inputenc}
\usepackage{amsmath,amssymb,booktabs,graphicx,multirow}
\usepackage[sort,compress]{cite}
\usepackage{xcolor}
\usepackage[verbose,hypertexnames=false]{hyperref}
\hypersetup{colorlinks=false,allbordercolors=blue,pdfborderstyle={/S/U/W 1}}
\usepackage{subcaption}   
\newcommand{\fitpanel}[2]{%
  \begin{subfigure}[b]{0.49\linewidth}
    \centering
    \includegraphics[width=\linewidth,trim={0 0 2.0cm 0.8cm},clip]{figures/#1.png}
    \caption{#2}
  \end{subfigure}}

\begin{document}
\markboth{Máté Csanád, Yan Huang, Dániel Kincses, Márton I. Nagy, Barnabás Pórfy}{Comparison of $\chi^2$ Variants for Poisson-Distributed Data and Ratios}

%
\catchline{}{}{}{}{}
%

\title{Comparison of $\chi^2$ Variants for Poisson-Distributed Data and Ratios}
\author{Máté Csanád, Yan Huang, Dániel Kincses, Márton I. Nagy, Barnabás Pórfy}
\address{ELTE Eötvös Loránd University, Institute of Physics and Astronomy,
Pázmány Péter sétány 1/a, Budapest, H-1117, Hungary}

\maketitle

\begin{history}
\received{Day Month Year}
\revised{Day Month Year}
\accepted{Day Month Year}
\published{Day Month Year}
\end{history}

\begin{abstract}
Fits to binned, Poisson-distributed data are common in high-energy and heavy-ion physics, and they are particularly delicate in correlation function measurements, where the fitted observable is the ratio of two histograms. We compare several goodness-of-fit estimators on large toy data sets: the Neyman and Pearson $\chi^2$, their Yates continuity-corrected versions, a Neyman $\chi^2$ with the variance shifted by $1/2$, the Poisson log-likelihood, and the correlation function likelihood in which both the signal and the reference histogram are treated as Poisson distributed. For a single histogram with mean occupancy $\lambda$, the Neyman $\chi^2$ underestimates the bin content by approximately one count, while the Pearson $\chi^2$ overestimates it by approximately half a count; shifting the variance by $1/2$ changes the Neyman result only at order $1/\lambda$, and only the log-likelihood recovers the mean without bias. For the ratio of two histograms, Neyman-type estimators are biased by approximately $-3/\lambda$ in relative terms (3\% at $\lambda=100$), whereas the Pearson and likelihood-based fits are unbiased in the symmetric configuration studied here. The Yates correction leaves the fitted values essentially unchanged but systematically deflates the $\chi^2$, resulting in unrealistically high confidence levels. Simple analytic expressions are derived that reproduce all observed biases. Since these biases do not decrease with the number of bins, whereas the statistical uncertainties do, they can dominate the uncertainty of high-statistics, finely binned measurements. We therefore recommend likelihood-based fits, in particular the correlation function likelihood, for femtoscopic and similar ratio analyses.
\end{abstract}


\section{Introduction}
Histograms of event or particle counts are the basic data format of high-energy and heavy-ion physics: particle spectra, invariant-mass distributions, efficiency and acceptance corrections, and correlation functions are all extracted from binned counts, and the physical parameters are obtained by fitting models to them. A prominent example is femtoscopy, where the two-particle correlation function is measured as the ratio $C(q) = A(q)/B(q)$ of a same-event pair distribution $A$ and a mixed-event reference distribution $B$~\cite{Lisa:2005dd}, and source parameters such as radii, correlation strengths, or L\'evy exponents are determined from fits to this ratio. Modern data sets are large, but analyses are also increasingly differential (in pair transverse momentum, centrality, multiplicity, or in three-dimensional relative momentum), so that many bins, in particular those at low relative momentum where the physics signal is concentrated, contain only moderate numbers of counts. In this regime the choice of the fit estimator is not a technicality. A bias of order one count per bin does not average out when more bins are added, while the statistical uncertainty of the fitted parameters decreases as the inverse square root of the total number of counts. Estimator-induced biases therefore become relatively more important as data sets grow, and they can easily exceed the statistical uncertainties in high-statistics measurements.

When fitting histograms where bin contents represent Poisson-distributed counts, the choice of the $\chi^2$ estimator significantly impacts the accuracy of the fitted parameters and the resulting goodness-of-fit statistics. This project implements and tests several $\chi^2$ definitions to evaluate their performance, particularly regarding the Neyman bias.

The Poisson distribution is fundamentally characterized by its asymmetry, particularly at lower mean values. This asymmetry is quantified through its higher-order moments: the skewness is given by $S=1/\sqrt{\lambda}$ and the excess kurtosis by $\kappa = 1/\lambda$.\footnote{In this paper, we use $\lambda$ for the Poisson parameter, or in other words, occupancy. Since this manuscript is related to the Workshop on Particle Correlations and Femtoscopy, it is important not to mix this $\lambda$ with the correlation function strength parameter usually denoted with the same Greek letter in femtoscopy.}

Because a Gaussian distribution is symmetric ($S = 0$ and $\kappa = 0$), approximating Poisson counts with a Gaussian-based $\chi^2$ is only an approximation, which improves as $\lambda$ increases. The dominant source of the Neyman bias~\cite{Contribution:112803}, however, is not the skewness itself but the fact that the Neyman $\chi^2$ estimates the variance from the observed count: bins that fluctuate downward are assigned a smaller variance and hence a larger weight in the fit, pulling the fitted values downward. This effect is of second order in the relative fluctuations, so the resulting relative bias scales as $1/\lambda$ (like the excess kurtosis), rather than as $1/\sqrt{\lambda}$ (like the skewness); see Sec.~\ref{s:results}. The variance shift by $0.5$ investigated in Sec.~\ref{ss:corr} is motivated by the best Gaussian approximation of the Poisson distribution; we test whether it reduces the bias.

\section{Estimating data-theory deviations}
Two primary scenarios are considered: fitting a single histogram and fitting the ratio of two histograms.

\subsection{Histogram fitting}
For an observed count $O_i$ and an expected theoretical value $E_i$ in the $i$-th bin, the general $\chi^2$ contribution is defined as:
\begin{equation}
    \chi^2 = \sum_i \frac{(O_i - E_i)^2}{\sigma_i^2}\label{e:chi2def}
\end{equation}
where $\sigma_i^2$ is the estimated variance, traditionally expressed, based on the Poisson distribution as
\begin{align}
\sigma^2_i = O_i.\label{e:sigmadef}
\end{align}

\subsection{Ratio fitting}
When fitting a ratio $C_i = A_i/B_i$ of two Poisson-distributed histograms, let $F_i$ be the fitted value of the ratio. The expected count in the numerator is then $F_i \cdot B_i$, which is compared against the observed count $A_i$. In some cases, an additional normalizing factor $N$ is used, hence the expected count is $N\cdot F_i \cdot B_i$. We leave this out for now, i.e. work with $N=1$.

 The variance in this case is traditionally estimated as
\begin{align}
\sigma^2_{C,i} = C_i^2\left(\frac{1}{A_i}+\frac{1}{B_i}\right) 
 = \frac{A_i}{B_i^2}+\frac{A_i^2}{B_i^3} = \frac{A_i}{B_i^2}\left(1 + \frac{A_i}{B_i}\right).\label{e:sigma_ratio}
\end{align}
Since $(C_i - F_i)^2/\sigma^2_{C,i} = (A_i - F_i B_i)^2/(B_i^2\sigma^2_{C,i})$, the corresponding variance of the numerator-space deviation $A_i - F_i B_i$ is $B_i^2\sigma^2_{C,i} = A_i\left(1 + A_i/B_i\right)$, which is the form used below.

\section{Variants for $\chi^2$}

\subsection{Neyman's $\chi^2$}\label{ss:neyman}
As outlined above in Eqs.~\eqref{e:chi2def}-\eqref{e:sigmadef}, the classical approach uses the observed counts to estimate variance:
\begin{equation}
    \chi^2_{\text{Neyman}} = \sum_i \frac{(O_i - E_i)^2}{O_i}\label{e:default}
\end{equation}
For fitting a ratio of histograms $A_i/B_i$ with a fitted value $F_i$, the equivalent expression is, based on Eq.~\eqref{e:sigma_ratio}:
\begin{equation}
    \chi^2_{\text{Neyman, ratio}} = \sum_i \frac{(A_i - F_i \cdot B_i)^2}{A_i \cdot \left(1 + \frac{A_i}{B_i}\right)}\label{e:default_ratio}
\end{equation}
This method is known to result in an underestimation of bin contents, the phenomenon already mentioned above, referred to as Neyman bias~\cite{Contribution:112803}.

\subsection{Pearson's $\chi^2$}\label{ss:pearson}
Pearson's variant~\cite{Pearson01071900} uses the expected value $E_i$ for the variance, which removes the correlation between the fluctuation and its variance estimate. As shown in Sec.~\ref{ss:histres}, this does not make the fit unbiased: for a single histogram it replaces the downward Neyman bias with an upward bias of approximately half a count.
\begin{equation}
    \chi^2_{\text{Pearson}} = \sum_i \frac{(O_i - E_i)^2}{E_i}\label{e:Pearson}
\end{equation}
For a ratio fit, the expression is implemented as:
\begin{equation}
    \chi^2_{\text{Pearson, ratio}} = \sum_i \frac{(A_i - F_i \cdot B_i)^2}{F_i \cdot B_i \cdot \left(1 + \frac{A_i}{B_i}\right)}\label{e:Pearson_ratio}
\end{equation}

\subsection{Yates' Continuity Correction}\label{ss:yates}
To improve the approximation of a discrete Poisson distribution by a continuous Gaussian, 0.5 is subtracted from the magnitude of the difference in the numerator~\cite{roth23646,Hitchcock2009}:
\begin{equation}
    \chi^2_{\text{Yates}} = \sum_i \frac{(|O_i - E_i| - 0.5)^2}{O_i},\label{e:Yates}
\end{equation}
while for a ratio fit, this becomes:
\begin{equation}
    \chi^2_{\text{Yates, ratio}} = \sum_i \frac{(|A_i - F_i \cdot B_i| - 0.5)^2}{A_i \cdot \left(1 + \frac{A_i}{B_i}\right)}\label{e:Yates_ratio}
\end{equation}
where the contribution is set to zero if $|O_i - E_i| < 0.5$ or $|A_i - F_i \cdot B_i| < 0.5$, respectively. 

\subsection{Yates-Pearson $\chi^2$}\label{ss:yates-pearson}
To combine the Pearson's variant with the Yates' correction, one may utilize:
\begin{equation}
    \chi^2_{\text{Y-P}} = \sum_i \frac{(|O_i - E_i| - 0.5)^2}{E_i},\label{e:YatesMod}
\end{equation}
and a ratio fit, the combined expression is:
\begin{equation}
    \chi^2_{\text{Y-P, ratio}} = \sum_i \frac{(|A_i - F_i \cdot B_i| - 0.5)^2}{F_i \cdot B_i \cdot \left(1 + \frac{A_i}{B_i}\right)}\label{e:YatesMod_ratio}
\end{equation}
where the contribution is again set to zero if the absolute difference is less than 0.5.

\subsection{A new type of shifted variance}\label{ss:corr}
Let us investigate the Gaussian distribution that most closely approximates a Poisson distribution with parameter $\lambda$. Let us start from a Poisson distribution with a mean (and variance) of $\lambda$:
\begin{align}
    p_n = \frac{\lambda^n}{n!}e^{-\lambda}.
\end{align}
A continuous version of this is (with $\Gamma$ representing the Gamma function):
\begin{align}
    p(x) = \frac{\lambda^x}{\Gamma(x+1)}e^{-\lambda}.
\end{align}
Here the following approximation can be utilized:
\begin{align}
    \ln \Gamma(1+x)=\frac{1}{2}\ln(2\pi)+\frac{1}{2}\ln x + x\ln x - x + \frac{1}{12x} + \mathcal{O}\left(\frac{1}{x^2}\right),
\end{align}
thus
\begin{align}
    \ln p(x) \approx x\ln\lambda -\lambda - \frac{1}{2}\ln(2\pi)-\frac{1}{2}\ln x - x\ln x + x -\frac{1}{12x}.
\end{align}
In this approximation, the peak of the distribution can be estimated via finding the point $x_0$ where the derivative of the above expression is zero. The leading-order result is $x_0 \approx \lambda$, and the next-to-leading-order one is:
\begin{align}
    x_0 \approx \lambda - \frac{1}{2},
\end{align}
and hence
\begin{align}
    \lambda \approx x_0 + \frac{1}{2}.
\end{align}
Thus the Gaussian describing a Poisson can be expressed as:
\begin{align}
    p(x) = A\exp\left[-\frac{(x-x_0)^2}{2\left(x_0+\frac{1}{2}\right)}\right]
\end{align}
This derivation suggests that a Gaussian fits a Poisson distribution optimally if the variance is shifted by $1/2$. The resulting $\chi^2$ is:
\begin{equation}
    \chi^2_{\text{corr}} = \sum_i \frac{(O_i - E_i)^2}{O_i + 0.5},\label{e:corr}
\end{equation}
and for ratio fitting, this becomes
\begin{equation}
    \chi^2_{\text{corr, ratio}} = \sum_i \frac{(A_i - F_i \cdot B_i)^2}{A_i\left(1 + \frac{A_i}{B_i}\right) + 0.5},\label{e:corr_ratio}
\end{equation}
It is instructive to consider a general shift $c$ of the variance, $\sigma_i^2 = O_i + c$, for the fit of a constant $E$. Minimizing the $\chi^2$ yields $\hat E = 1/\langle 1/(O+c)\rangle - c$, and expanding around $\lambda$ with the Poisson moments gives
\begin{align}
\hat E \approx \lambda - 1 + \frac{c-1}{\lambda}.\label{e:shift_bias}
\end{align}
The leading-order bias of $-1$ count is therefore independent of the shift, and $c$ only affects the subleading $\mathcal{O}(1/\lambda)$ term: $c=0$ (Neyman) gives $\lambda - 1 - 1/\lambda$ and $c=0.5$ gives $\lambda - 1 - 0.5/\lambda$, in agreement with the results of Table~\ref{tab:hist_results}. Removing the leading bias requires shifting the numerator as well, as done, e.g., in the $\chi^2_\gamma$ statistic of Mighell~\cite{Mighell:1999}, $\sum_i (O_i + \min(O_i,1) - E_i)^2/(O_i+1)$, for which $\hat E = \lambda/(1-e^{-\lambda})$.

\subsection{Log-Likelihood (LL)}\label{ss:ll}
To avoid Gaussian assumptions entirely, log-likelihood optimization is used~\cite{BAKER1984437}. The equivalent $\chi^2$ definition for Poisson statistics is:
\begin{equation}
    \chi^2_{\text{LL}} = \sum_i 2 \left[ E_i - O_i + O_i \ln\left(\frac{O_i}{E_i}\right) \right],\label{e:LL}
\end{equation}
and for histogram ratios, it becomes:
\begin{equation}
    \chi^2_{\text{LL,ratio}} = \sum_i \left[ F_{i}\cdot B_i - A_{i} + A_{i} \ln\left(\frac{A_{i}}{F_i\cdot B_{i}}\right) \right].\label{e:LL_ratio}
\end{equation}

Note that Eq.~\eqref{e:LL_ratio} treats $B_i$ as exactly known, i.e., it neglects the fluctuations of the reference histogram. The fitted value is then $\hat F = \sum_i A_i / \sum_i B_i$, which is unbiased, but the value of $\chi^2_{\text{LL,ratio}}$ is a reliable goodness-of-fit measure only if $B_i$ has much higher statistics than $A_i$.

Note furthermore that in our implementation the conventional factor of 2 is omitted. Since
$\mathrm{Var}(A_i - F_i B_i) = F_i B_i (1+F_i)$, this happens to account
for the fluctuations of $B_i$ when $F_i \approx 1$ and $A$ and $B$ have equal
statistics, as in our test; the same coincidence makes the MINUIT~\cite{James:1975dr} uncertainty
(obtained with error definition 1) correct. In general, Eq.~\eqref{e:LL_ratio}
with the factor 2 is appropriate only if $B_i$ has much higher statistics than
$A_i$, otherwise the CF likelihood of Eq.~\eqref{e:cf_ll} should be used.

\subsection{Correlation Function Likelihood (CF Likelihood)}\label{ss:cfl}

In correlation function analysis, both the signal distribution $A_i$ and the reference distribution $B_i$ are treated as independent Poisson processes. For this case, beyond what is mentioned in the above subsection, another  log-likelihood (equivalently defined as $-2\ln\mathcal{L}$) is constructed as~\cite{E802:2002fwe}:

\begin{equation}
\chi^2_{\text{CF-LL}} =
-2 \sum_i \left[
A_i \ln\left(\frac{F_i (A_i + B_i)}{A_i (F_i+1)}\right)
+ B_i \ln\left(\frac{A_i + B_i}{B_i (F_i+1)}\right)
\right],
\label{e:cf_ll}
\end{equation}
where, for consistency with the previous subsections, $F_i$ denotes the fitted value of the correlation function (i.e., $F_i$ takes the role of $C$ in Ref.~\cite{E802:2002fwe}, and it must not be confused with the measured ratio $C_i = A_i/B_i$).

This expression corresponds exactly to the joint Poisson likelihood of two statistically independent measurements under a multiplicative correlation hypothesis, after profiling out the underlying nuisance expectation values.

\section{Numerical results and analysis}\label{s:results}

The various $\chi^2$ definitions were tested using a histogram with $N_{\text{bins}} = 400\,000$ filled with $N_{\text{hits}} = 40$M, 80M, and 160M uniformly distributed random entries, corresponding to mean bin occupancies of $\lambda = N_{\text{hits}}/N_{\text{bins}} = 100$, 200, and 400. For the ratio fits, two statistically independent histograms $A$ and $B$ were generated in the same way, with the same number of entries each. In all cases a constant was fitted, hence $\text{NDF} = N_{\text{bins}} - 1 = 399\,999$. The results for the histogram constant fit and the ratio fit are summarized below.

All fits were performed within the ROOT framework~\cite{Brun:1997pa}, using the MINUIT minimizer~\cite{James:1975dr} through the \texttt{TVirtualFitter} interface: for each estimator, the corresponding expression was implemented as a user-defined objective function, minimized with MIGRAD, and the parameter uncertainty was obtained from the second derivative of the objective function at its minimum (HESSE), with error definition~1 (see Sec.~\ref{ss:ll} for the implications of this choice for the ratio log-likelihood). The macros used for the tests are publicly available~\cite{PoissonChi2Test}. The confidence levels quoted below were computed from the minimum value of the objective function and the NDF, assuming the asymptotic $\chi^2$ distribution. For each value of $\lambda$, a single data set was generated (one histogram for the single-histogram test, and one pair of histograms $A$ and $B$ for the ratio test), and all estimators were applied to the same data set. The quoted values thus reflect the differences between the estimators, not independent statistical fluctuations. The histograms were filled with a fixed total number of entries; the bin contents are therefore multinomially distributed, which for $N_{\text{bins}} = 400\,000$ is indistinguishable from independent Poisson distributions in each bin, but it implies that the sum of all bin contents equals $N_{\text{hits}}$ exactly. Consequently, the log-likelihood fit of a single histogram, whose minimum is the sample mean, returns exactly $N_{\text{hits}}/N_{\text{bins}}$, and in the ratio test $\sum_i A_i = \sum_i B_i$ holds exactly. Bins with zero content were excluded wherever the estimator is undefined for them; at the occupancies studied here, no such bins occur.

When interpreting the goodness-of-fit results, note that for $\text{NDF} \approx 4\times10^5$ the standard deviation of $\chi^2/\text{NDF}$ is $\sqrt{2/\text{NDF}} \approx 0.0022$. Differences of a few times $10^{-3}$ in $\chi^2/\text{NDF}$, and the corresponding differences in confidence levels, between single realizations are thus not significant; systematic effects are visible only if they are much larger than this.

\subsection{Histogram Fitting Results}\label{ss:histres}
For a histogram where the expected value is $\lambda = N_{\text{hits}}/N_{\text{bins}}$, the performance of different estimators is shown in Table \ref{tab:hist_results}.

\begin{table}[tbp]
\centering
\caption{Histogram fitting results for a $N_{\text{bins}} = 400\,000$ histogram. $\lambda$ denotes the bin occupancy, i.e., the mean number of hits per bin.}
\label{tab:hist_results}
\begin{tabular}{l c c c c c c}
\hline
Case & $N_{\text{hits}}$ & $\lambda$ & Fit & $\chi^2$ & $\chi^2$/NDF & Prob (C.L.) \\
\hline
\multirow{3}{*}{\shortstack[l]{Neyman\\(Sec.~\ref{ss:neyman})}}
 & 40M  & 100 & $98.991\pm0.016$  & 403752.4 & 1.0094 & $<0.01$\% \\
 & 80M  & 200 & $198.992\pm0.022$ & 403065.3 & 1.0077 & 0.03\% \\
 & 160M & 400 & $398.995\pm0.032$ & 401828.2 & 1.0046 & 2.05\% \\
\hline
\multirow{3}{*}{\shortstack[l]{Pearson\\(Sec.~\ref{ss:pearson})}}
 & 40M  & 100 & $100.498\pm0.016$ & 398785.9 & 0.9970 & 91.26\% \\
 & 80M  & 200 & $200.501\pm0.022$ & 400412.5 & 1.0010 & 32.17\% \\
 & 160M & 400 & $400.501\pm0.032$ & 400526.1 & 1.0013 & 27.77\% \\
\hline
\multirow{3}{*}{\shortstack[l]{Yates\\(Sec.~\ref{ss:yates})}}
 & 40M  & 100 & $98.998\pm0.016$  & 372521.7 & 0.9313 & 100.00\% \\
 & 80M  & 200 & $198.997\pm0.023$ & 380866.0 & 0.9522 & 100.00\% \\
 & 160M & 400 & $398.998\pm0.032$ & 386072.6 & 0.9652 & 100.00\% \\
\hline
\multirow{3}{*}{\shortstack[l]{Yates \&\\Pearson\\(Sec.~\ref{ss:yates-pearson})}}
 & 40M  & 100 & $100.486\pm0.016$ & 367947.6 & 0.9199 & 100.00\% \\
 & 80M  & 200 & $200.492\pm0.023$ & 378357.4 & 0.9459 & 100.00\% \\
 & 160M & 400 & $400.494\pm0.032$ & 384823.0 & 0.9621 & 100.00\% \\
\hline
\multirow{3}{*}{\shortstack[l]{Shifted\\variance\\(Sec.~\ref{ss:corr})}}
 & 40M  & 100 & $98.996\pm0.016$  & 401681.7 & 1.0042 & 3.01\% \\
 & 80M  & 200 & $198.995\pm0.022$ & 402044.4 & 1.0051 & 1.12\% \\
 & 160M & 400 & $398.997\pm0.032$ & 401322.7 & 1.0033 & 6.96\% \\
\hline
\multirow{3}{*}{\shortstack[l]{Likelihood\\(Sec.~\ref{ss:ll})}}
 & 40M  & 100 & $100.000\pm0.016$ & 400406.0 & 1.0010 & 32.43\% \\
 & 80M  & 200 & $200.000\pm0.022$ & 401284.4 & 1.0032 & 7.54\% \\
 & 160M & 400 & $400.000\pm0.032$ & 400956.0 & 1.0024 & 14.23\% \\
\hline
\end{tabular}
\end{table}

The Default (Neyman) method underestimates the constant by approximately one count at all occupancies, i.e., by about 60 times the statistical uncertainty at $\lambda=100$. The shifted-variance variant of Sec.~\ref{ss:corr} changes the fitted value only by $0.002$--$0.005$, as predicted by Eq.~\eqref{e:shift_bias}; its apparent improvement in the confidence level is within the statistical fluctuation of $\chi^2/\text{NDF}$ discussed above. The Pearson fit is biased in the opposite direction: minimizing Eq.~\eqref{e:Pearson} for a constant gives $\hat E = \sqrt{\langle O^2\rangle} = \sqrt{\lambda^2+\lambda} \approx \lambda + 1/2 - 1/(8\lambda)$, which reproduces the observed $100.498$, $200.501$ and $400.501$. The Yates correction leaves the fitted values of the respective uncorrected methods nearly unchanged. The only unbiased estimator is the log-likelihood, whose minimum for a constant is exactly the sample mean. Thus, in absolute terms, the Neyman and Pearson biases do not depend on $\lambda$ (about $-1$ and $+1/2$ counts, respectively), while the relative biases scale as $1/\lambda$.

\subsection{Ratio Fitting Results}\label{ss:ratiores}
The performance was also evaluated for the ratio of two histograms, where the expected value is $1.000$.

\begin{table}[tbp]
\centering
\caption{Ratio fitting results for a ratio of two $N_{\text{bins}} = 400\,000$ histograms. Here the expected value is always 1.0.}
\label{tab:ratio_results}
\begin{tabular}{l c c c c c}
\hline
Case & $N_{\text{hits}}$ & Fit & $\chi^2$ & $\chi^2$/NDF & Prob (C.L.) \\
\hline
\multirow{3}{*}{\shortstack[l]{Neyman\\(Sec.~\ref{ss:neyman})}}
 & 40M  & $0.9703\pm0.0002$ & 395655.8 & 0.9891 & 100.00\% \\
 & 80M  & $0.9851\pm0.0002$ & 396850.3 & 0.9921 & 99.98\% \\
 & 160M & $0.9925\pm0.0001$ & 398584.5 & 0.9965 & 94.32\% \\
\hline
\multirow{3}{*}{\shortstack[l]{Pearson\\(Sec.~\ref{ss:pearson})}}
 & 40M  & $1.0000\pm0.0002$ & 401110.9 & 1.0028 & 10.70\% \\
 & 80M  & $1.0000\pm0.0002$ & 399787.9 & 0.9995 & 59.30\% \\
 & 160M & $1.0000\pm0.0001$ & 399818.0 & 0.9995 & 57.99\% \\
\hline
\multirow{3}{*}{\shortstack[l]{Yates\\(Sec.~\ref{ss:yates})}}
 & 40M  & $0.9703\pm0.0002$ & 373538.2 & 0.9338 & 100.00\% \\
 & 80M  & $0.9851\pm0.0002$ & 381142.6 & 0.9529 & 100.00\% \\
 & 160M & $0.9925\pm0.0001$ & 387426.7 & 0.9686 & 100.00\% \\
\hline
\multirow{3}{*}{\shortstack[l]{Yates \&\\Pearson\\(Sec.~\ref{ss:yates-pearson})}}
 & 40M  & $0.9997\pm0.0002$ & 378948.0 & 0.9474 & 100.00\% \\
 & 80M  & $0.9999\pm0.0002$ & 384061.9 & 0.9602 & 100.00\% \\
 & 160M & $1.0000\pm0.0001$ & 388657.5 & 0.9716 & 100.00\% \\
\hline
\multirow{3}{*}{\shortstack[l]{Shifted\\variance\\(Sec.~\ref{ss:corr})}}
 & 40M  & $0.9704\pm0.0002$ & 394630.7 & 0.9866 & 100.00\% \\
 & 80M  & $0.9851\pm0.0002$ & 396345.8 & 0.9909 & 100.00\% \\
 & 160M & $0.9925\pm0.0001$ & 398333.2 & 0.9958 & 96.89\% \\
\hline
\multirow{3}{*}{\shortstack[l]{Likelihood\\(Sec.~\ref{ss:ll})}}
 & 40M  & $1.0000\pm0.0002$ & 403109.3 & 1.0078 & 0.03\% \\
 & 80M  & $1.0000\pm0.0002$ & 400796.9 & 1.0020 & 18.61\% \\
 & 160M & $1.0000\pm0.0001$ & 400279.8 & 1.0007 & 37.65\% \\
\hline
\multirow{3}{*}{\shortstack[l]{CF Likelihood\\(Sec.~\ref{ss:cfl})}}
 & 40M  & $1.0000\pm0.0002$ & 402131.6 & 1.0053 & 0.86\% \\
 & 80M  & $1.0000\pm0.0002$ & 400290.8 & 1.0007 & 37.19\% \\
 & 160M & $1.0000\pm0.0001$ & 400069.1 & 1.0002 & 46.85\% \\
\hline
\end{tabular}
\end{table}

In the ratio tests, the Pearson and likelihood methods again outperform the Default and corrected Neyman approaches (and the Yates-corrected Neyman version), which all exhibit a $3\%$ downward bias at 40M hits. The bias scales as $1/\lambda$: expanding Eq.~\eqref{e:default_ratio} to second order in the relative fluctuations of $A_i$ and $B_i$ at $F=1$ gives $\hat F \approx 1 - 3/\lambda$, i.e., $0.970$, $0.985$ and $0.9925$ for $\lambda = 100$, 200 and 400, in excellent agreement with Table~\ref{tab:ratio_results}. The relative bias is three times larger than for a single histogram, because the fluctuations of both histograms enter the variance estimate.

The unbiased result of the Pearson ratio fit should be interpreted with care. Minimizing Eq.~\eqref{e:Pearson_ratio} gives $\hat F^2 = \sum_i \frac{A_i^2}{A_i+B_i} \big/ \sum_i \frac{B_i^2}{A_i+B_i}$, which is symmetric under $A \leftrightarrow B$; hence $\hat F=1$ is guaranteed on average only when $A$ and $B$ have the same statistics and the true ratio is one, as in the present test. For log-likelihood fits the fitted value $\sum_i A_i/\sum_i B_i$ is unbiased in general, while the CF likelihood of Eq.~\eqref{e:cf_ll} also provides a correct goodness-of-fit measure, since it accounts for the fluctuations of both histograms.

\subsection{Overview of the results}\label{ss:overview}
\begin{figure}[tbp]
    \centering
    \includegraphics[width=\linewidth]{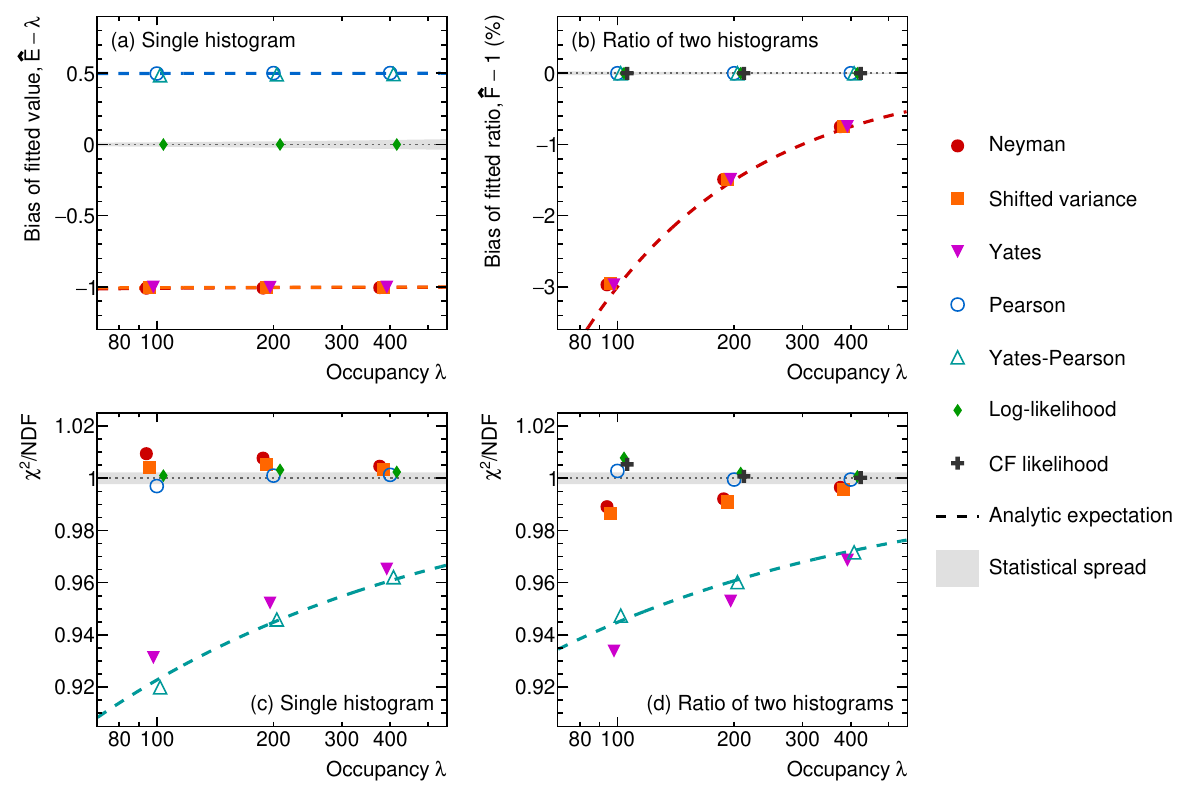}
    \caption{Summary of the estimator tests as a function of the bin occupancy $\lambda$. Top: bias of the fitted value for a single histogram (a), in counts, and for the ratio of two histograms (b), in percent. Bottom: $\chi^2/\text{NDF}$ at the minimum for a single histogram (c) and for the ratio (d). Markers show the fit results of Tables~\ref{tab:hist_results} and \ref{tab:ratio_results}, slightly shifted horizontally for visibility. Dashed lines show the analytic expectations: Eq.~\eqref{e:shift_bias} for the Neyman and shifted-variance fits, $\sqrt{\lambda^2+\lambda}-\lambda$ for the Pearson fit, $-3/\lambda$ for the Neyman ratio fit, and Eq.~\eqref{e:yates_chi2} with $\sigma^2=\lambda$ and $2\lambda$ for the Yates-Pearson $\chi^2$. Grey bands indicate the statistical uncertainty of the fitted value (top) and the expected spread $\pm\sqrt{2/\text{NDF}}$ of $\chi^2/\text{NDF}$ (bottom).}
    \label{fig:summary}
\end{figure}

The results of Tables~\ref{tab:hist_results} and \ref{tab:ratio_results} are summarized in Fig.~\ref{fig:summary}, together with the analytic expectations derived above; the individual fits are shown in~\ref{app:fits}. The top panels illustrate the magnitude and the scaling of the biases. For a single histogram (panel a), the estimators separate into three groups that do not depend on $\lambda$: the Neyman-type estimators (Neyman, shifted variance, and Yates) lie at about $-1$ count, the Pearson-type estimators (Pearson and Yates-Pearson) at about $+1/2$ count, and the log-likelihood at zero. The Neyman and shifted-variance results are indistinguishable on this scale, as expected from Eq.~\eqref{e:shift_bias}, and the analytic curves reproduce all points. The grey band, which indicates the statistical uncertainty of the fitted value, is narrow compared with the biases: the Neyman bias corresponds to about 60, 45 and 30 standard deviations at $\lambda = 100$, 200 and 400, respectively. For the ratio of two histograms (panel b), the Neyman-type estimators follow the expected $-3/\lambda$ dependence, while all other estimators are consistent with zero bias in this symmetric test configuration (cf.\ the discussion in Sec.~\ref{ss:ratiores}).

The bottom panels show the goodness-of-fit statistic. For the Pearson and likelihood-based estimators, $\chi^2/\text{NDF}$ is consistent with unity within a few times its expected spread $\sqrt{2/\text{NDF}}$, indicated by the grey band. The Neyman-type estimators tend to deviate from unity, by up to about five standard deviations, upward for a single histogram and downward for the ratio; since they are evaluated at a biased minimum and with a variance estimated from the data, their $\chi^2$ values are not reliable goodness-of-fit measures either. The strongest effect is seen for the estimators that include Yates' correction, which yield $\chi^2/\text{NDF}$ values well below unity and hence confidence levels of $100\%$. This can be quantified: for Gaussian fluctuations $\delta$ with variance $\sigma^2$, one has
\begin{align}
\frac{\langle(|\delta|-1/2)^2\rangle}{\sigma^2} \approx 1 - \sqrt{\frac{2}{\pi}}\frac{1}{\sigma} + \frac{1}{4\sigma^2}.\label{e:yates_chi2}
\end{align}
With $\sigma^2 = \lambda$ for a single histogram this gives $\chi^2/\text{NDF} \approx 0.923$, $0.945$ and $0.961$ for $\lambda = 100$, 200 and 400, and with $\sigma^2 \approx 2\lambda$ for the ratio it gives $0.945$, $0.961$ and $0.972$, in good agreement with the Yates-Pearson results, as shown by the dashed curves in panels (c) and (d). The Yates-corrected Neyman estimator lies above the curve in panel (c) and below it in panel (d), reflecting the corresponding shifts of the uncorrected Neyman $\chi^2$. Since the fitted values are essentially unaffected by the correction (cf.\ panels a and b), the Yates correction does not mitigate the bias; it only deflates the goodness-of-fit statistic. Note that the relative deflation, of order $1/\sqrt{\lambda}$, decreases more slowly with the occupancy than the biases of the fitted values, which are of order $1/\lambda$.

\section{Summary}\label{s:summary}
We have compared several $\chi^2$ and likelihood-based estimators for fits to Poisson-distributed histograms and to ratios of such histograms, as encountered in correlation function measurements. Using toy data with $4\times10^5$ bins and mean occupancies of $\lambda = 100$--$400$, we find that estimators which take the variance from the observed counts (the Neyman $\chi^2$ and its variants) systematically underestimate the fitted quantity. For a single histogram the bias is about one count, independent of $\lambda$; for a ratio of two equal-statistics histograms the relative bias is about $3/\lambda$. These values are reproduced by simple analytic expansions, which also show that the dominant mechanism is the correlation between the observed count and its variance estimate. Accordingly, shifting the variance by $1/2$, as suggested by the best Gaussian approximation of the Poisson distribution, only modifies the bias at order $1/\lambda$ and does not remove it. The Pearson $\chi^2$ is not unbiased either: for a single histogram it overestimates the bin content by about half a count, and its unbiased behaviour in the ratio test follows from the symmetry of the test configuration. The Yates continuity correction leaves the fitted values essentially unchanged, but it deflates $\chi^2$ by a predictable amount of relative order $1/\sqrt{\lambda}$, leading to confidence levels close to $100\%$ that do not reflect the true fit quality. The Poisson log-likelihood is the only estimator that is unbiased in all tested cases, and for histogram ratios the correlation function likelihood, which treats both histograms as Poisson distributed, additionally yields a well-behaved goodness-of-fit statistic.

Because these biases do not shrink as the number of bins grows, whereas the statistical uncertainties do, they can exceed the statistical errors in high-statistics, multi-differential analyses. We therefore recommend likelihood-based fits, and for correlation functions the CF likelihood of Eq.~\eqref{e:cf_ll}, as the default choice. Natural extensions of this study include low occupancies ($\lambda \lesssim 20$, including empty bins), reference histograms with higher statistics than the signal (as typical for mixed-event backgrounds), non-flat correlation functions and free normalization parameters, as well as the evaluation of confidence levels from ensembles of toy experiments rather than from single realizations.

\section*{Acknowledgments}
This work was supported by the NKFIH grants PD-146589 and NKKP 152097.

\section*{ORCID}
\noindent Máté Csanád - \url{https://orcid.org/0000-0002-3154-6925}\\
\noindent Yan Huang - \url{https://orcid.org/0000-0002-1499-6051}\\
\noindent Dániel Kincses - \url{https://orcid.org/0000-0002-8880-6732}\\
\noindent Márton I. Nagy - \url{https://orcid.org/0000-0001-7887-681X}\\
\noindent Barnabás Pórfy - \url{https://orcid.org/0000-0001-5724-9737}

\bibliographystyle{ws-ijmpa}
\bibliography{ref}

\appendix
\section{Individual fit results}\label{app:fits}
The individual fits underlying Tables~\ref{tab:hist_results} and \ref{tab:ratio_results} and Fig.~\ref{fig:summary} are shown in Figs.~\ref{fig:histogram} and \ref{fig:ratio}, for the single-histogram and the ratio tests, respectively. Each figure spans three pages, one for each occupancy ($\lambda = 100$, 200 and 400, corresponding to $N_{\text{hits}} = 40$M, 80M and 160M), and each page shows the same data set fitted with the different estimators. The fitted constant is indicated by the red line, and the fit results are displayed in each panel. For the ratio test, the CF likelihood of Eq.~\eqref{e:cf_ll} is shown; the results of the ratio log-likelihood of Eq.~\eqref{e:LL_ratio} are listed in Table~\ref{tab:ratio_results}.

\begin{figure}[p]
    \centering
    \fitpanel{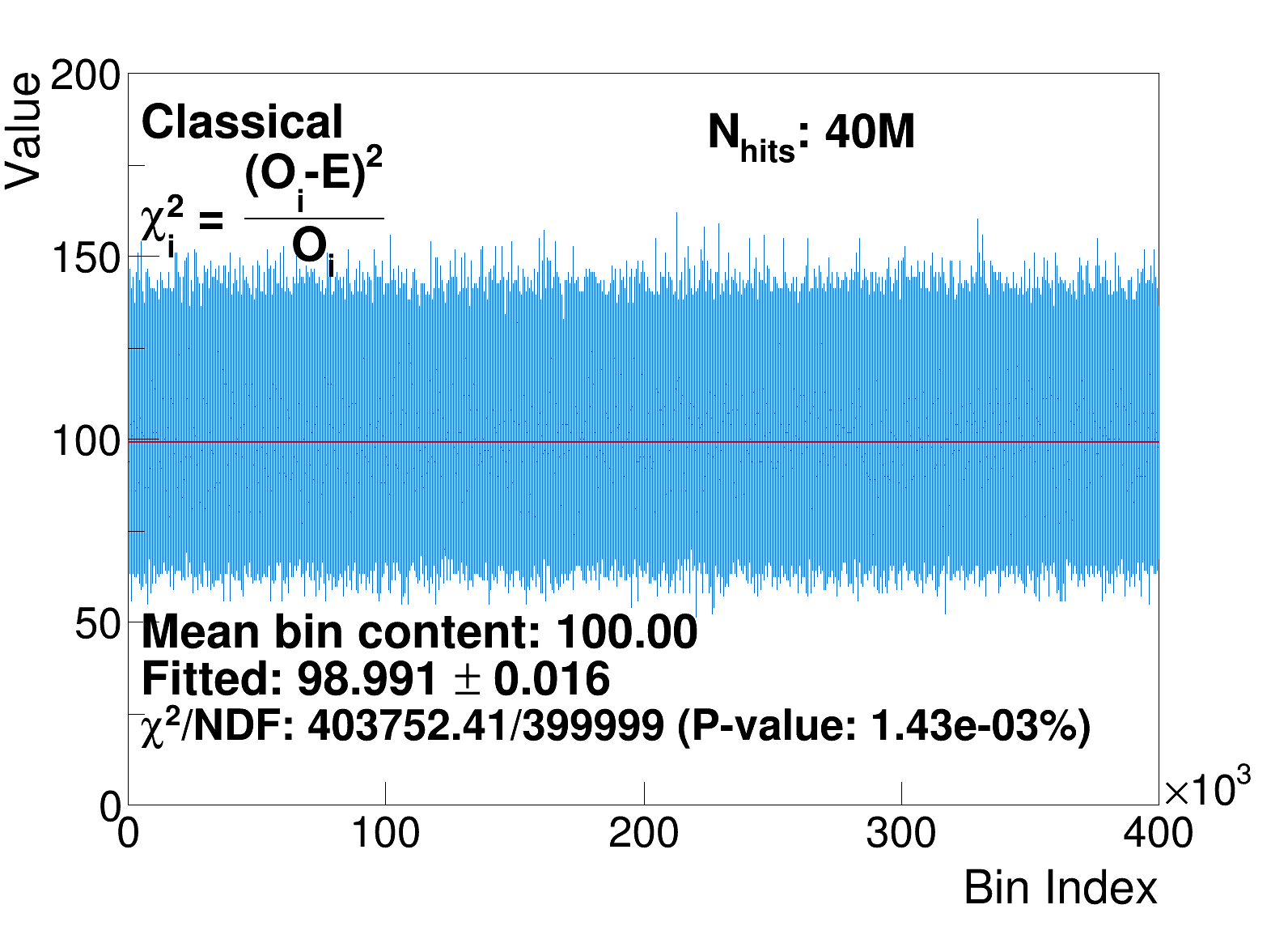}{Neyman, Eq.~\eqref{e:default}}\hfill
    \fitpanel{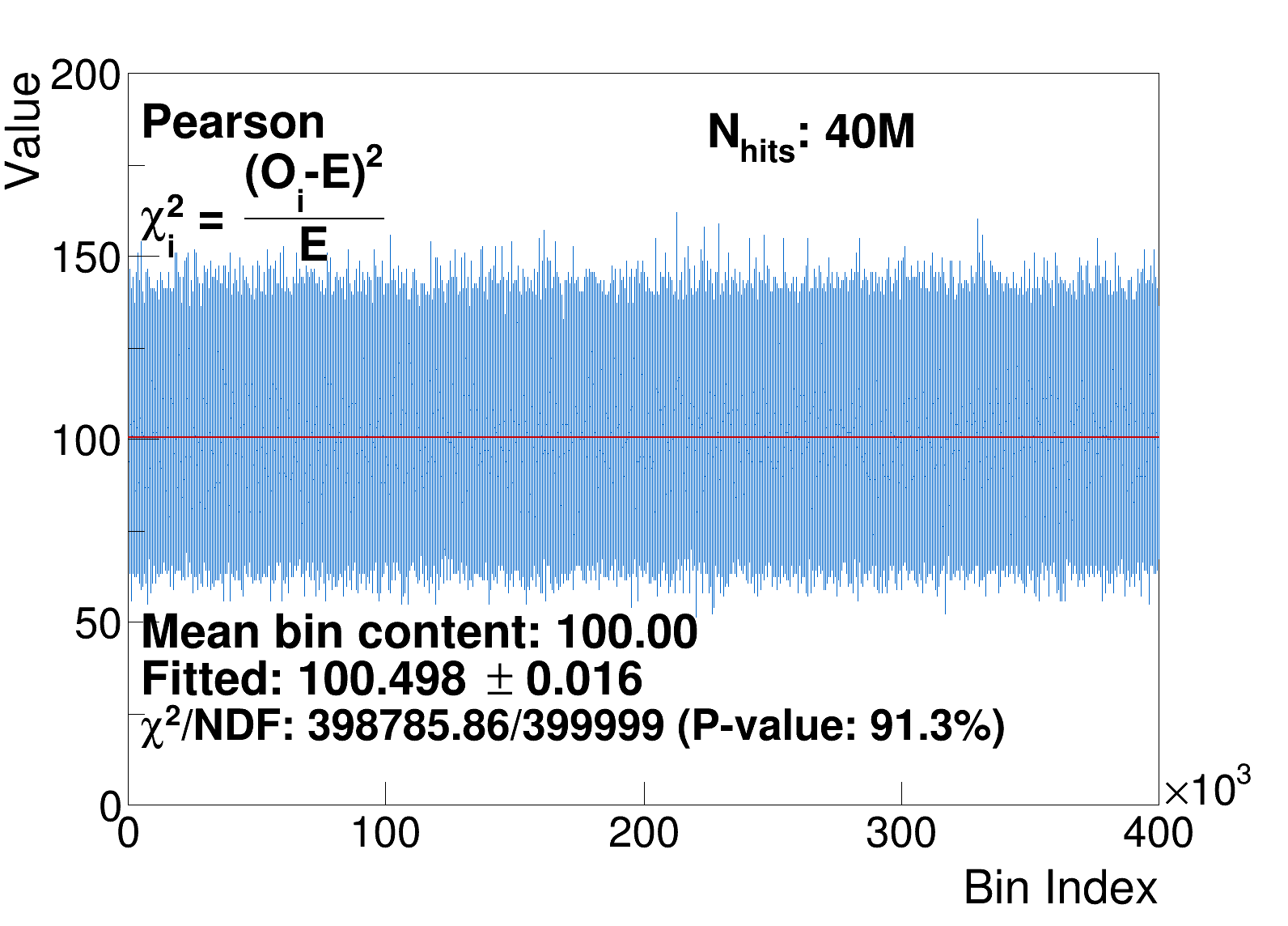}{Pearson, Eq.~\eqref{e:Pearson}}\\[1ex]
    \fitpanel{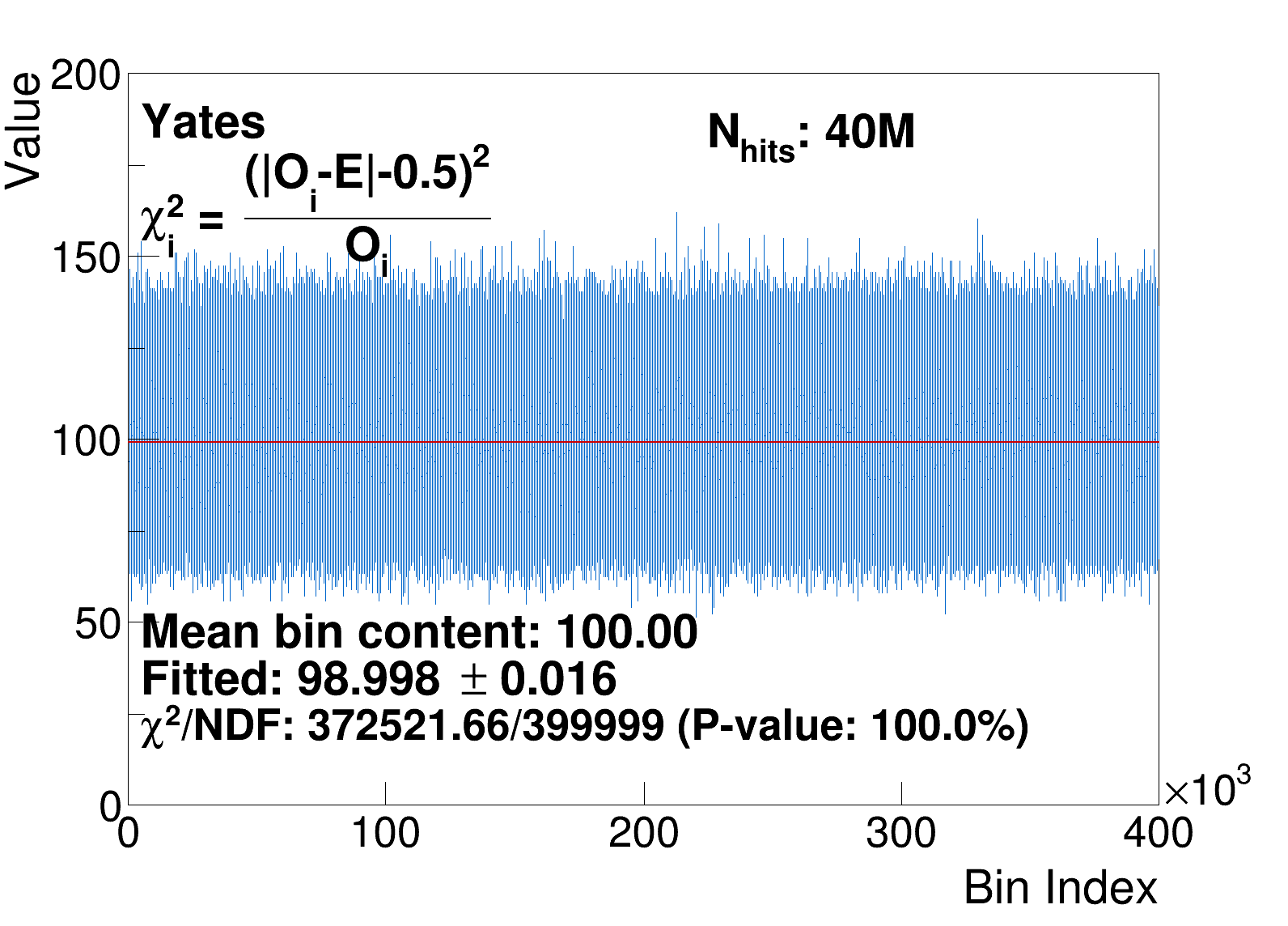}{Yates, Eq.~\eqref{e:Yates}}\hfill
    \fitpanel{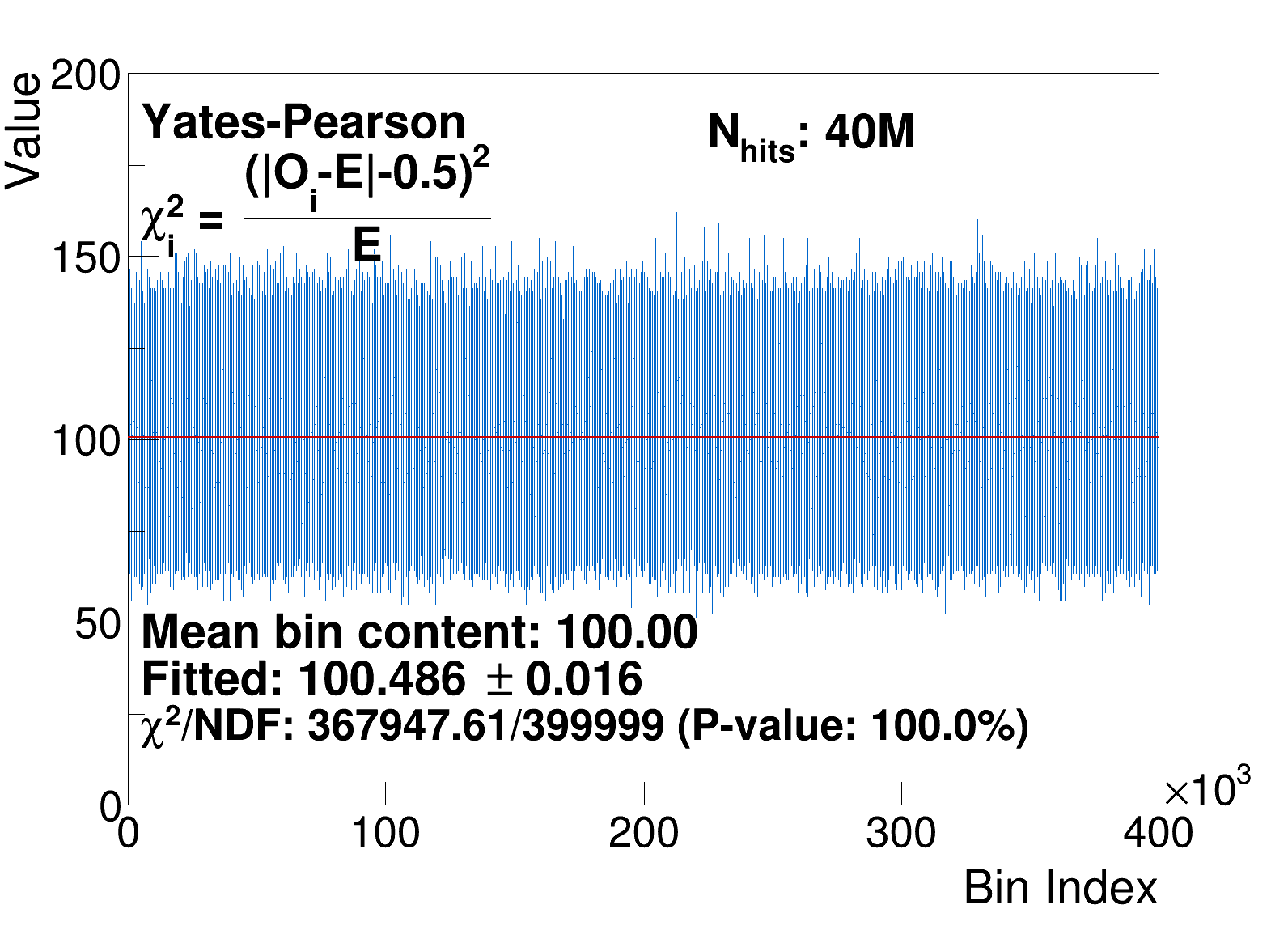}{Yates-Pearson, Eq.~\eqref{e:YatesMod}}\\[1ex]
    \fitpanel{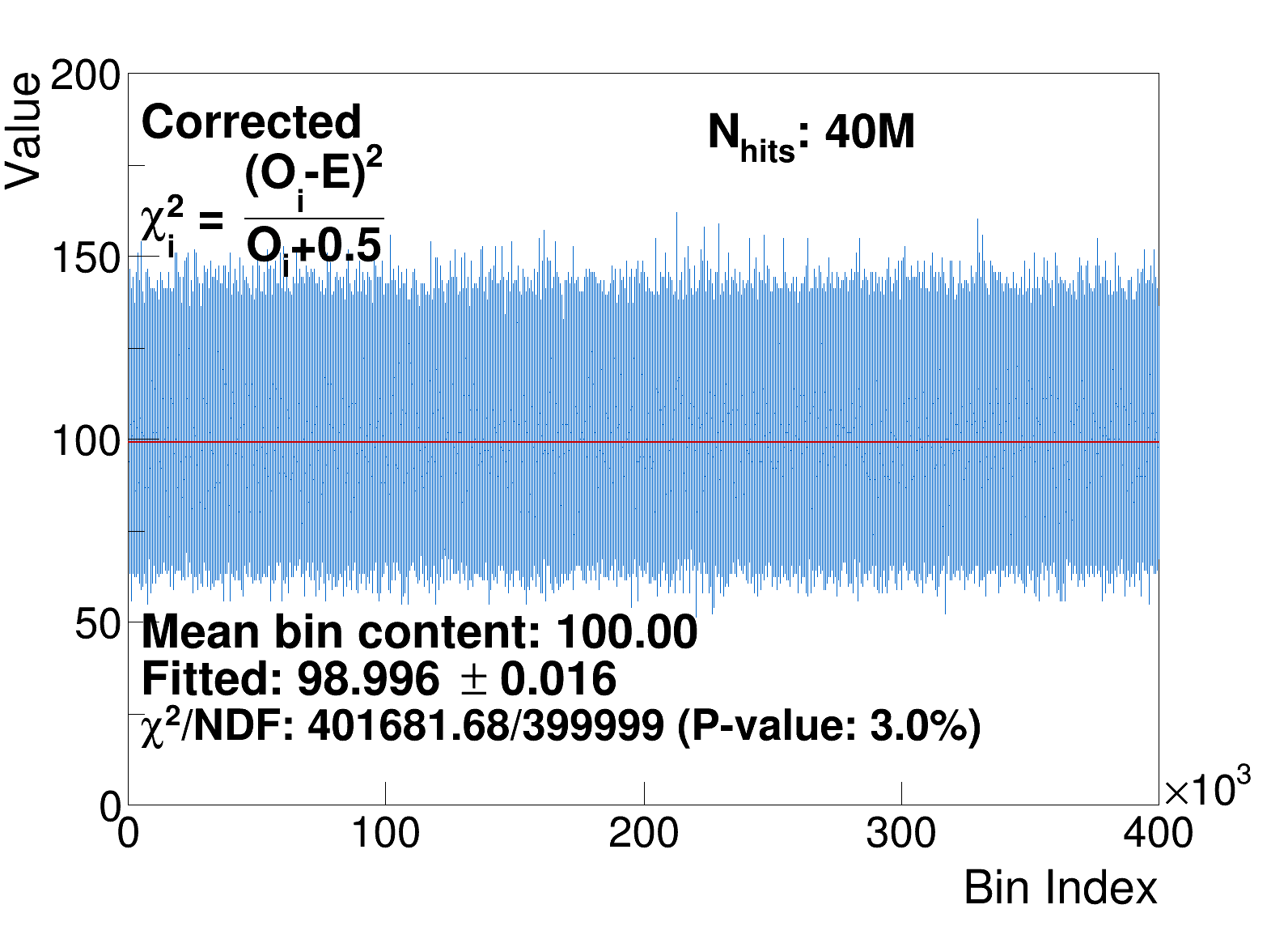}{Shifted variance, Eq.~\eqref{e:corr}}\hfill
    \fitpanel{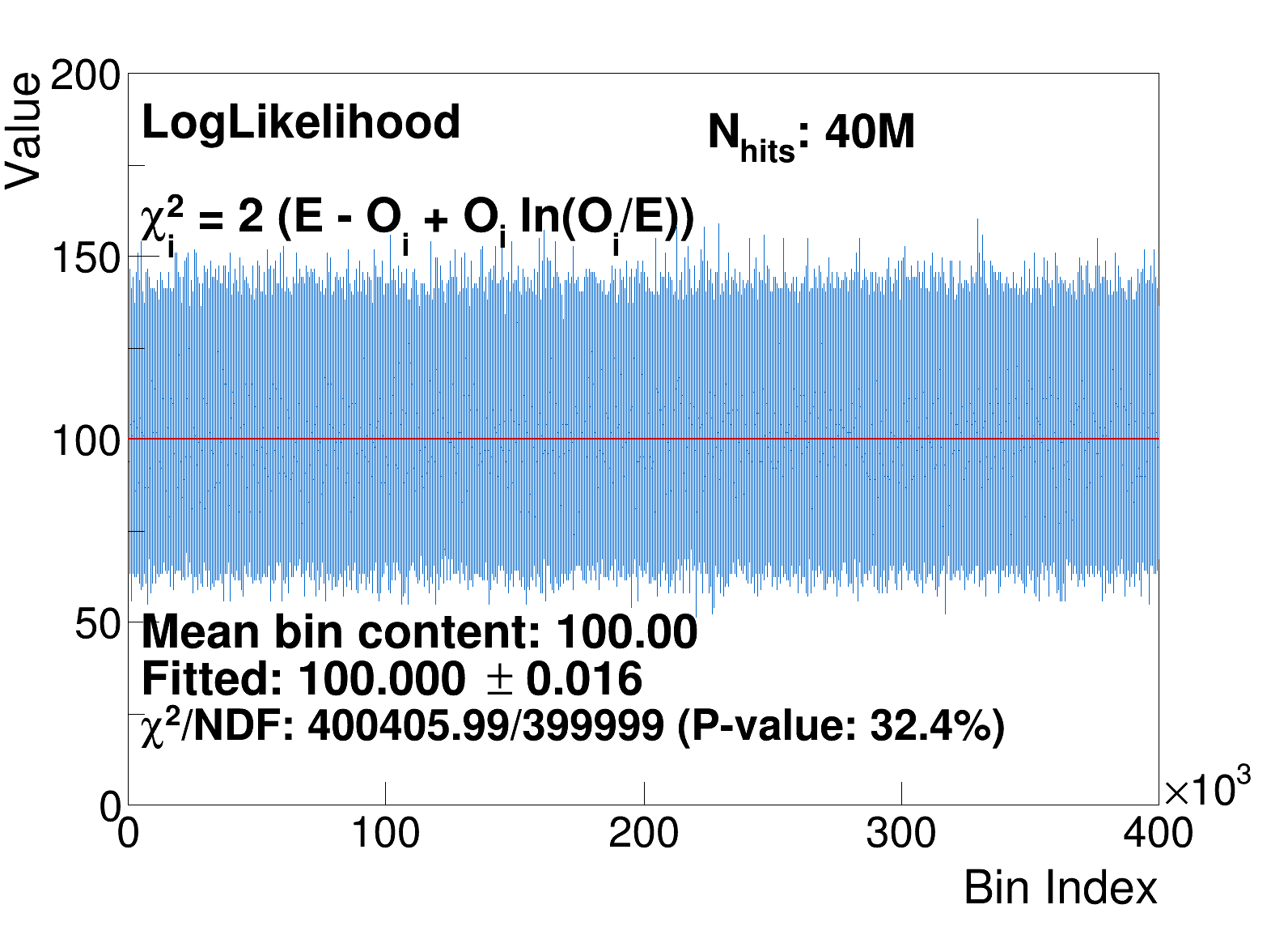}{Log-likelihood, Eq.~\eqref{e:LL}}
    \caption{Fits of a constant to a histogram with $N_{\text{bins}} = 400\,000$ bins, filled with uniformly distributed random entries, using the estimators of Sec.~\ref{ss:neyman}--\ref{ss:ll}. All panels on a page show the same data set. This page: $N_{\text{hits}} = 40$M ($\lambda = 100$).}
    \label{fig:histogram}
\end{figure}

\begin{figure}[p]
    \ContinuedFloat
    \centering
    \fitpanel{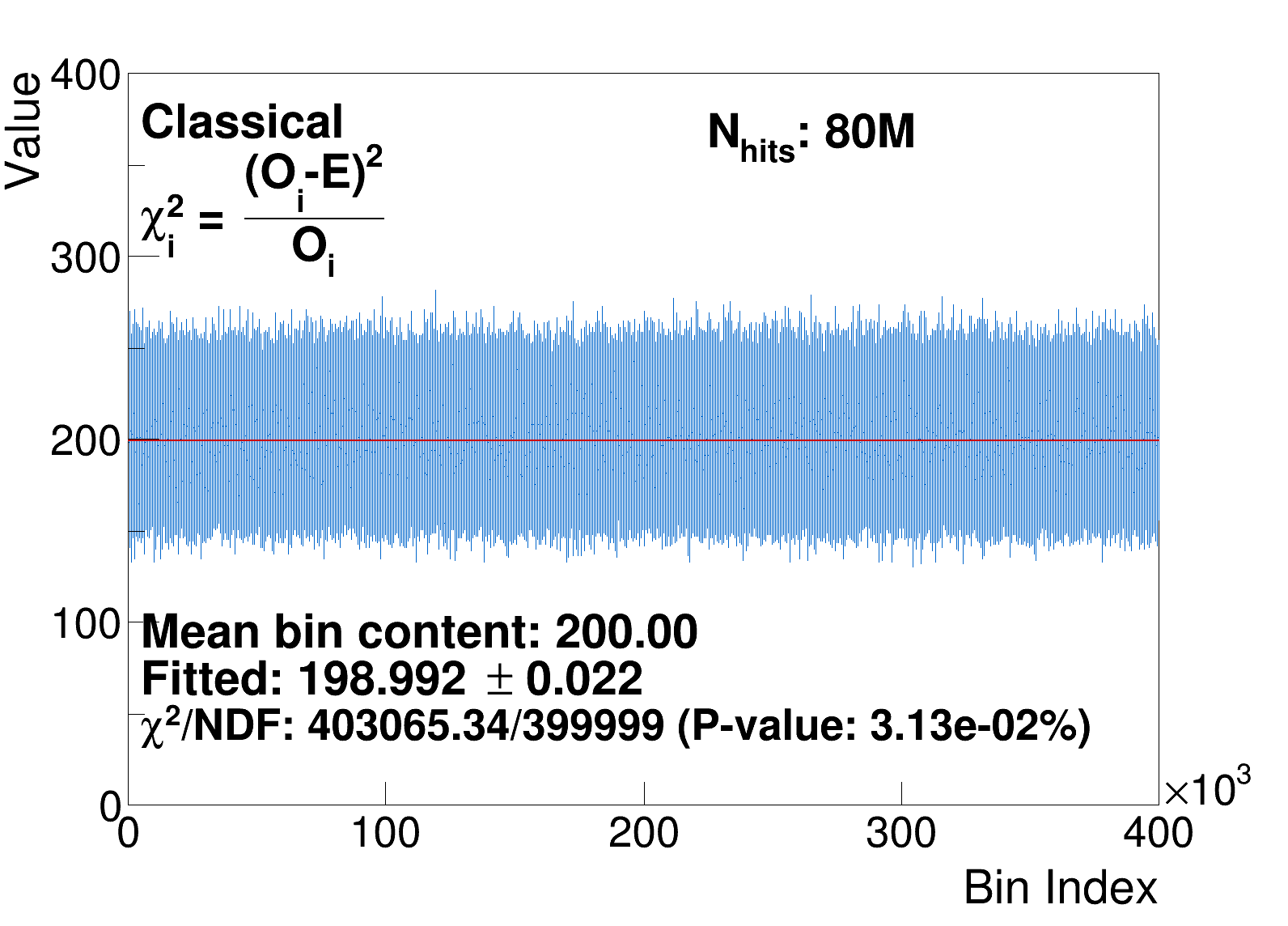}{Neyman, Eq.~\eqref{e:default}}\hfill
    \fitpanel{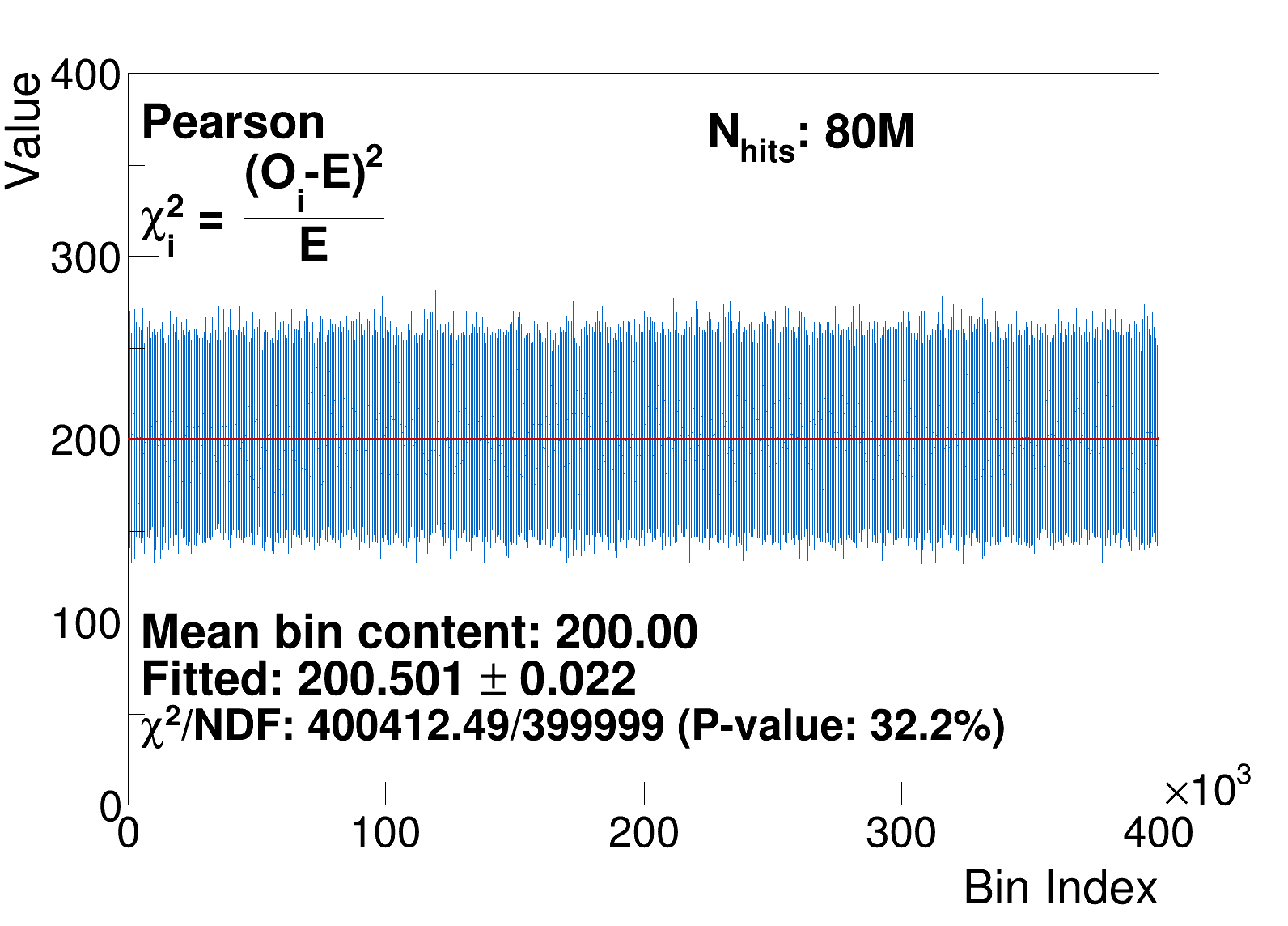}{Pearson, Eq.~\eqref{e:Pearson}}\\[1ex]
    \fitpanel{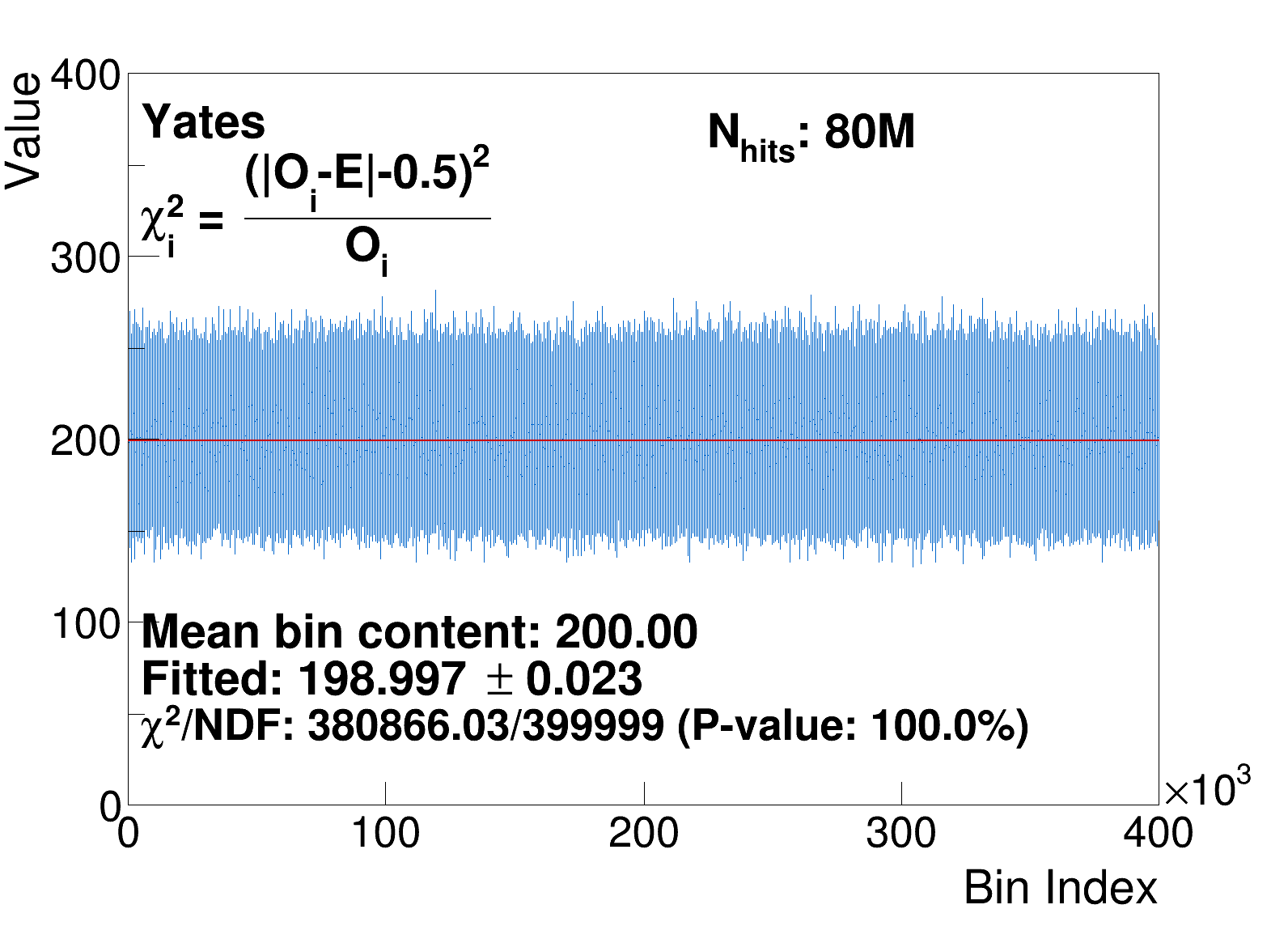}{Yates, Eq.~\eqref{e:Yates}}\hfill
    \fitpanel{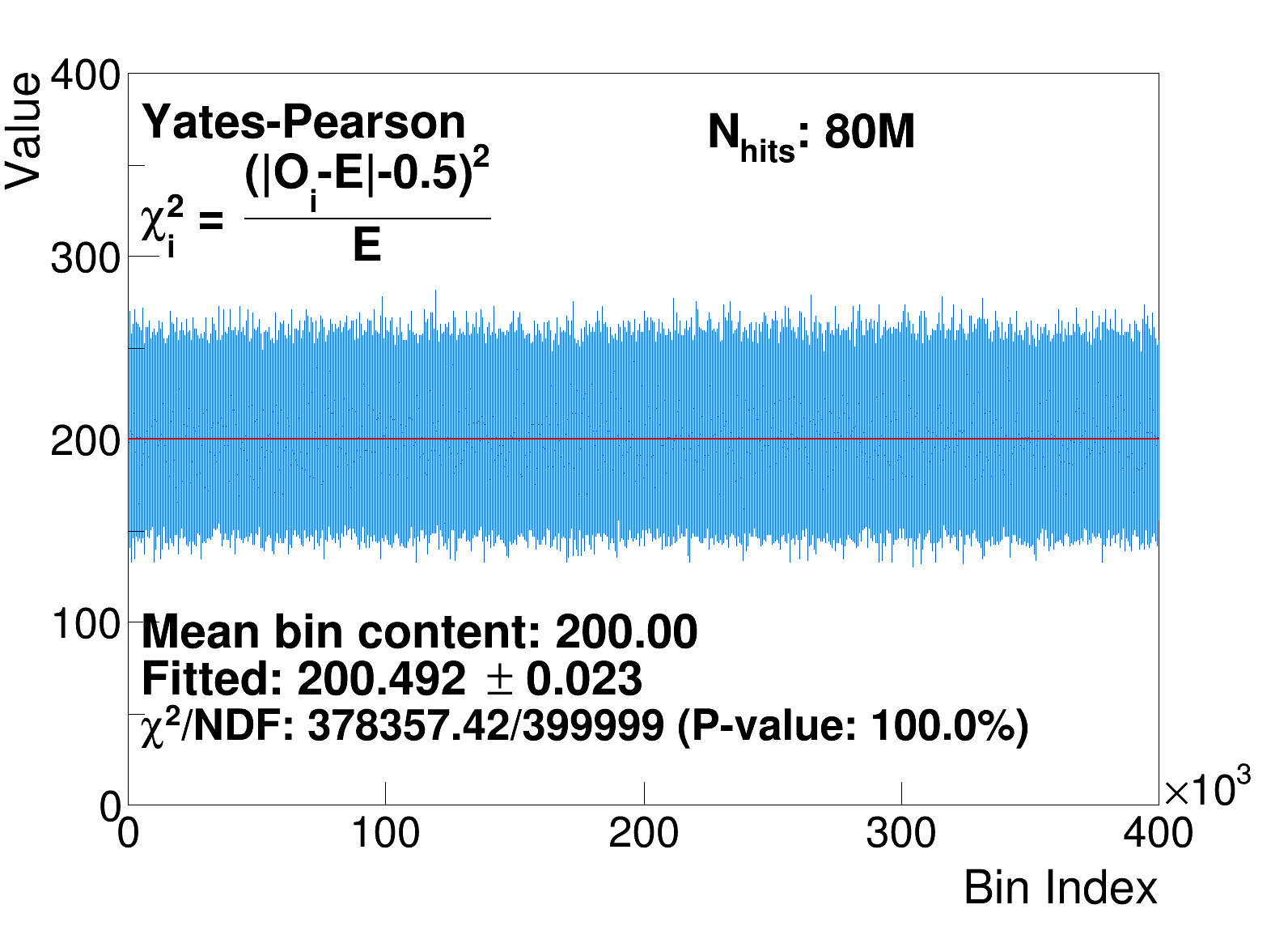}{Yates-Pearson, Eq.~\eqref{e:YatesMod}}\\[1ex]
    \fitpanel{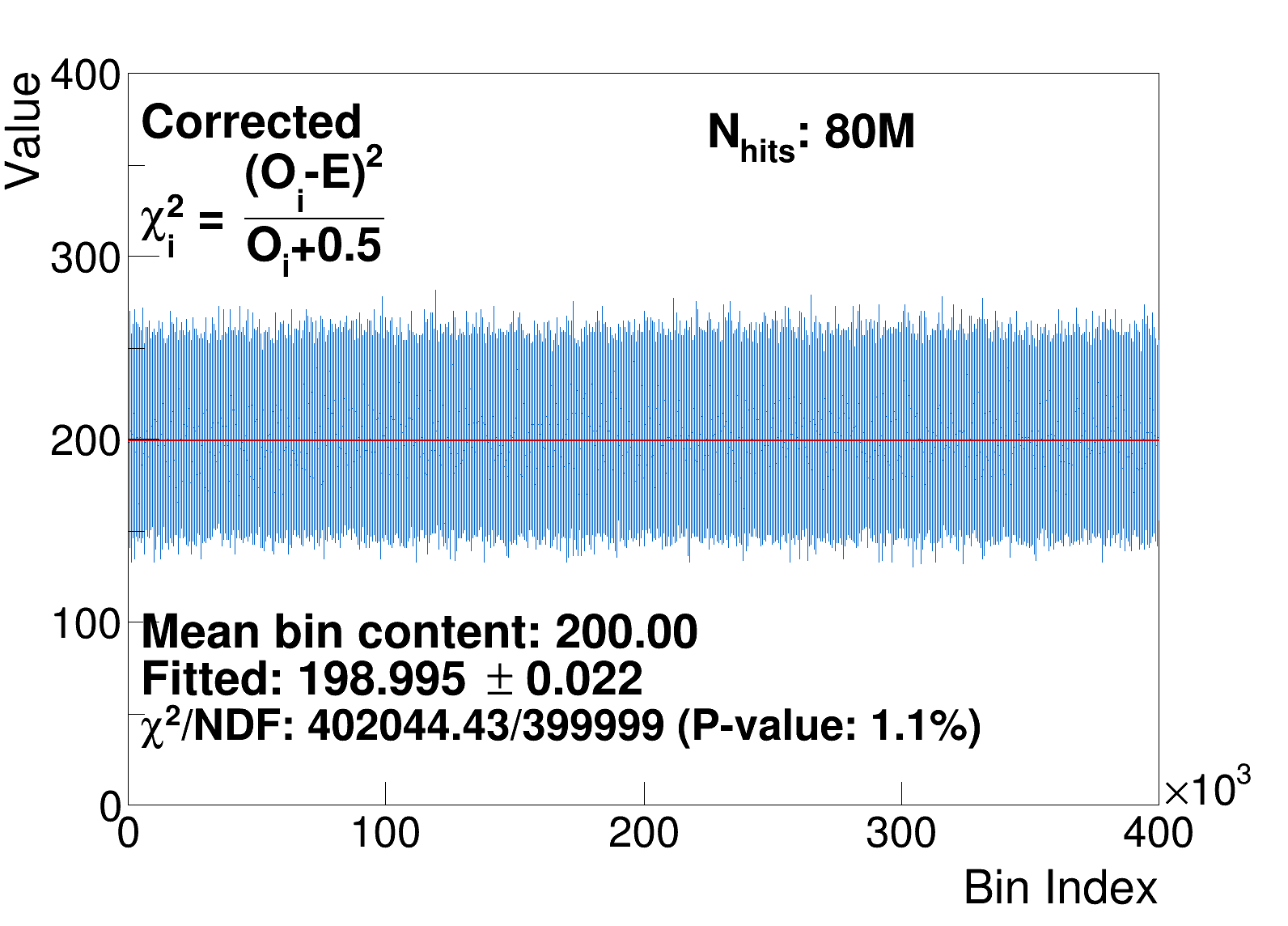}{Shifted variance, Eq.~\eqref{e:corr}}\hfill
    \fitpanel{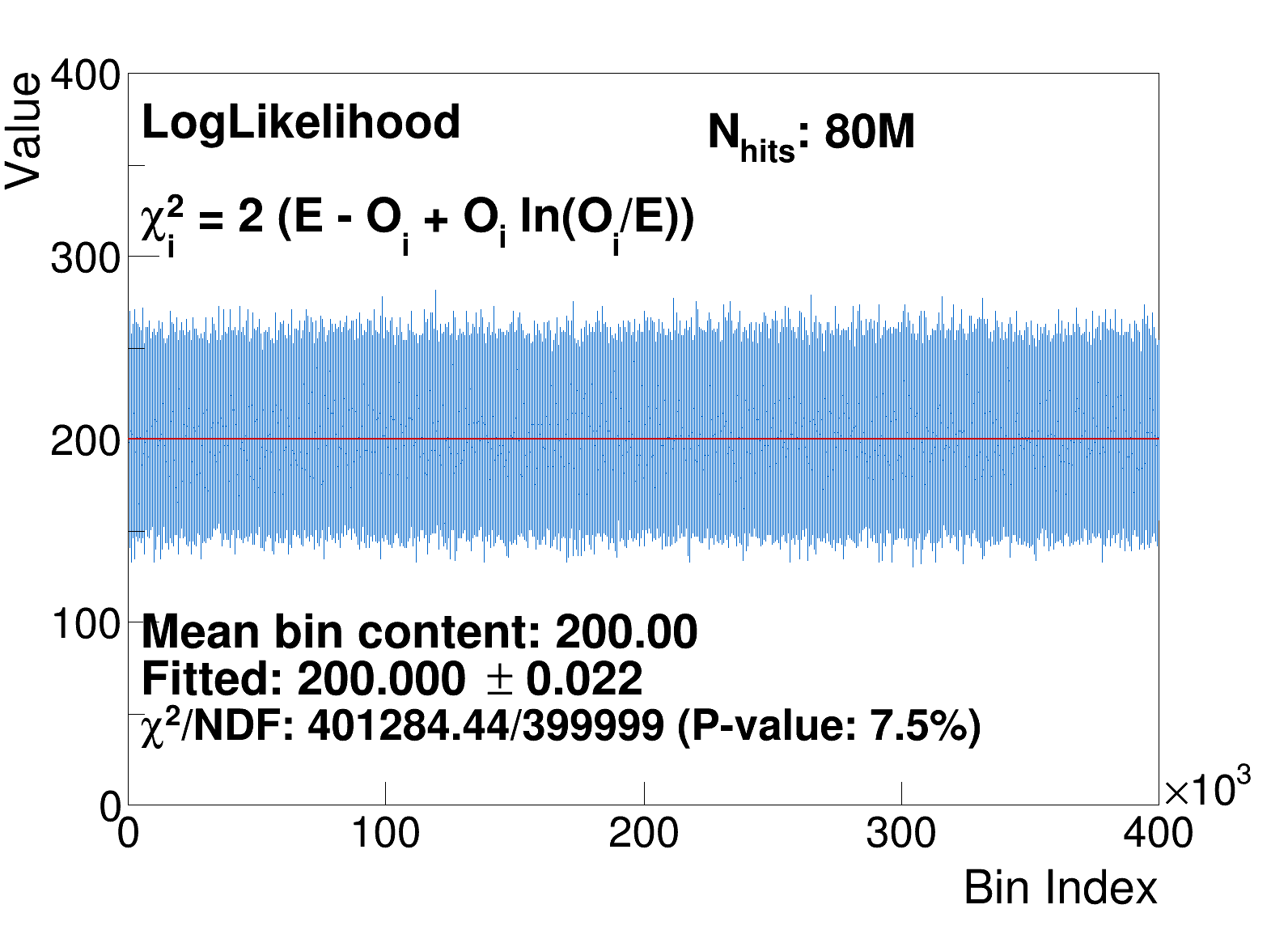}{Log-likelihood, Eq.~\eqref{e:LL}}
    \caption[]{(Continued.) $N_{\text{hits}} = 80$M ($\lambda = 200$).}
\end{figure}

\begin{figure}[p]
    \ContinuedFloat
    \centering
    \fitpanel{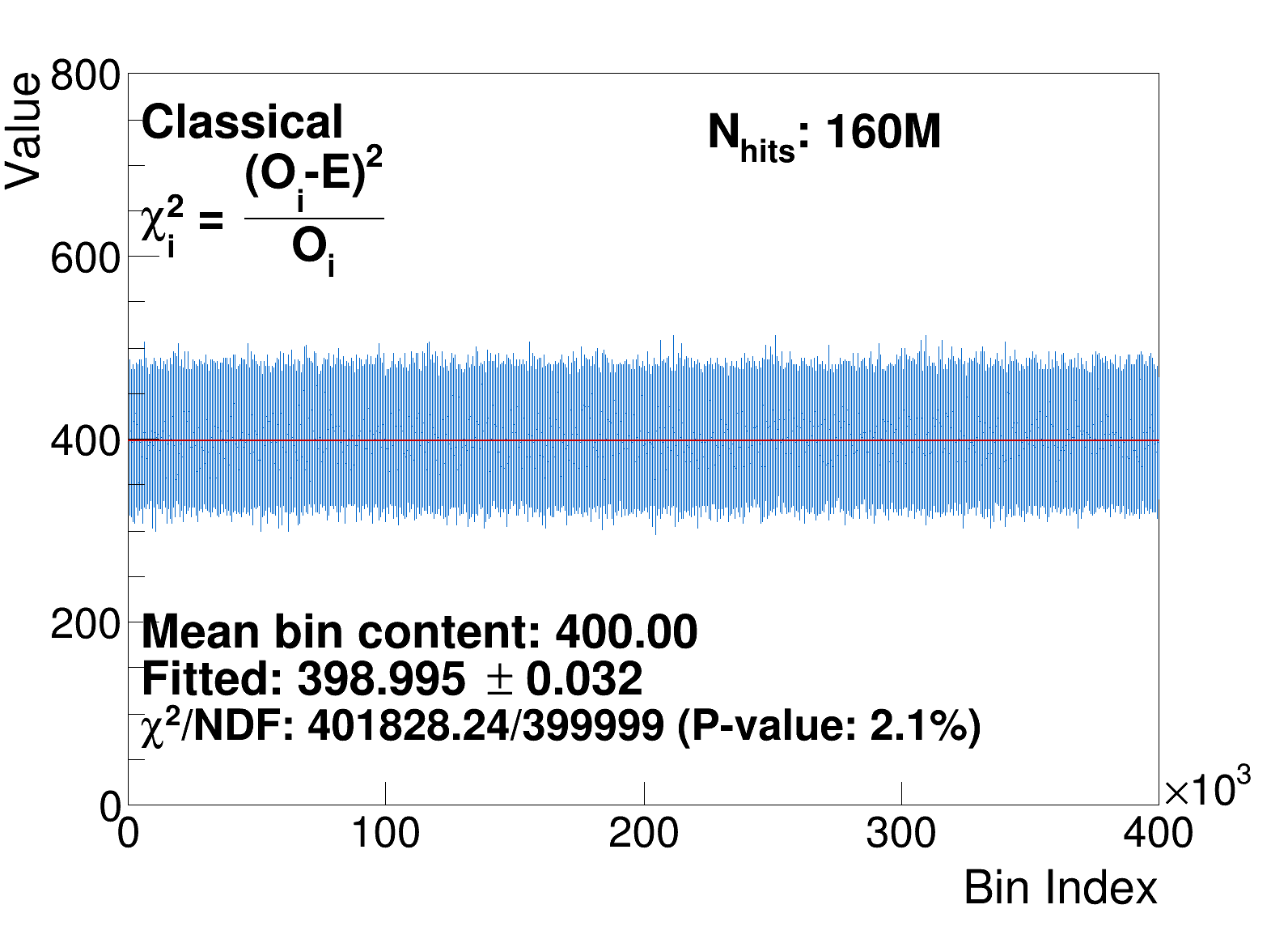}{Neyman, Eq.~\eqref{e:default}}\hfill
    \fitpanel{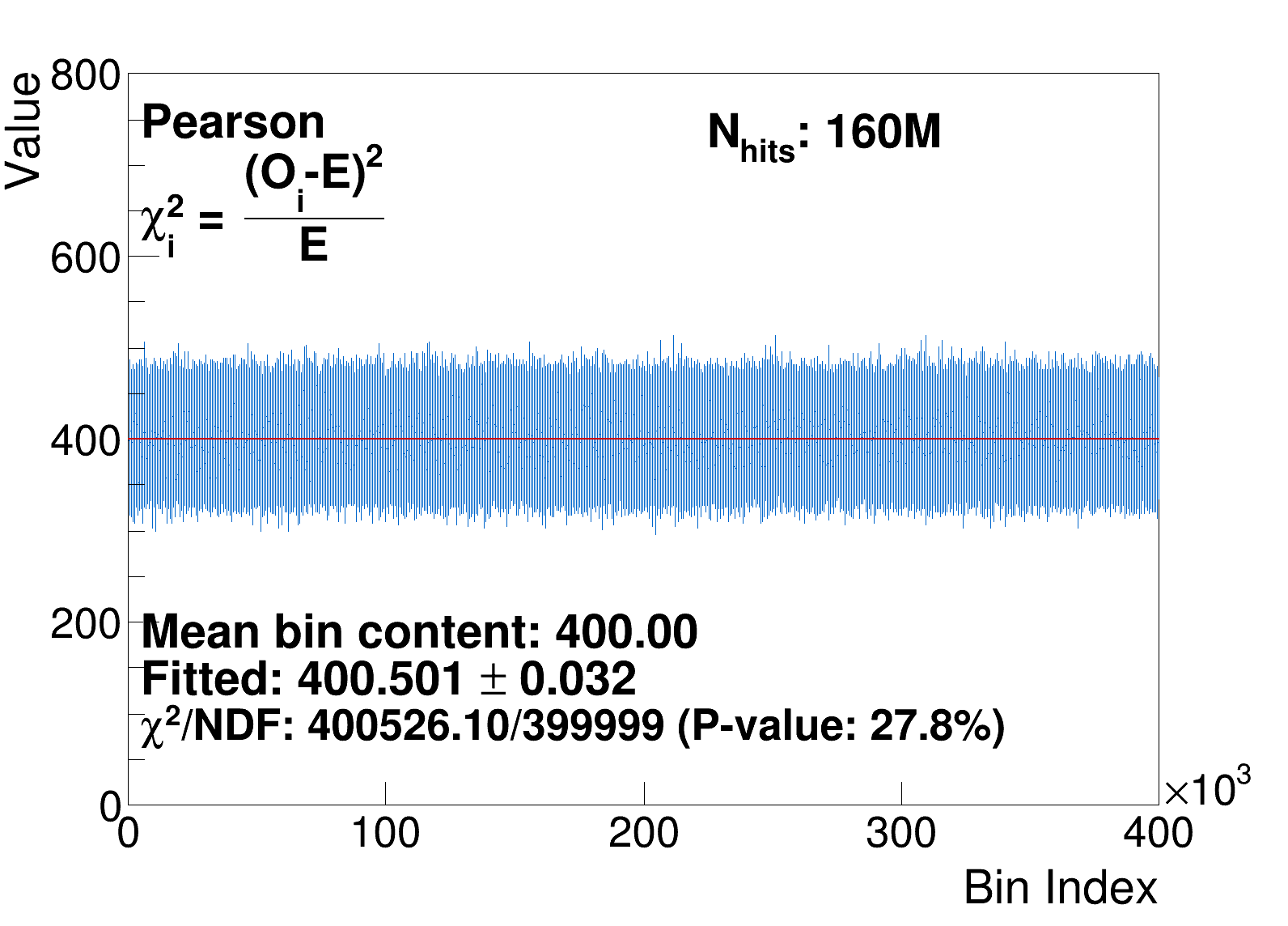}{Pearson, Eq.~\eqref{e:Pearson}}\\[1ex]
    \fitpanel{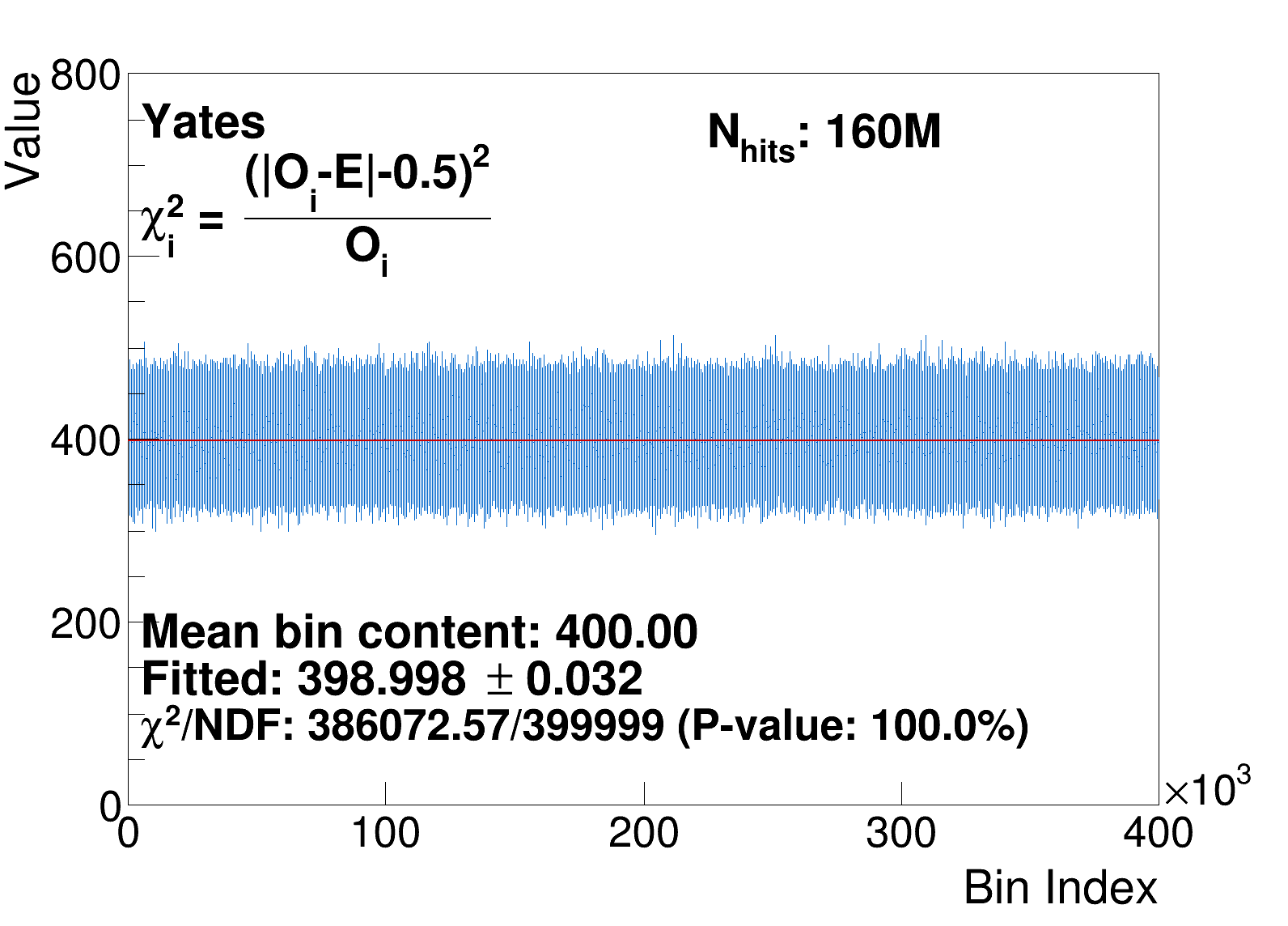}{Yates, Eq.~\eqref{e:Yates}}\hfill
    \fitpanel{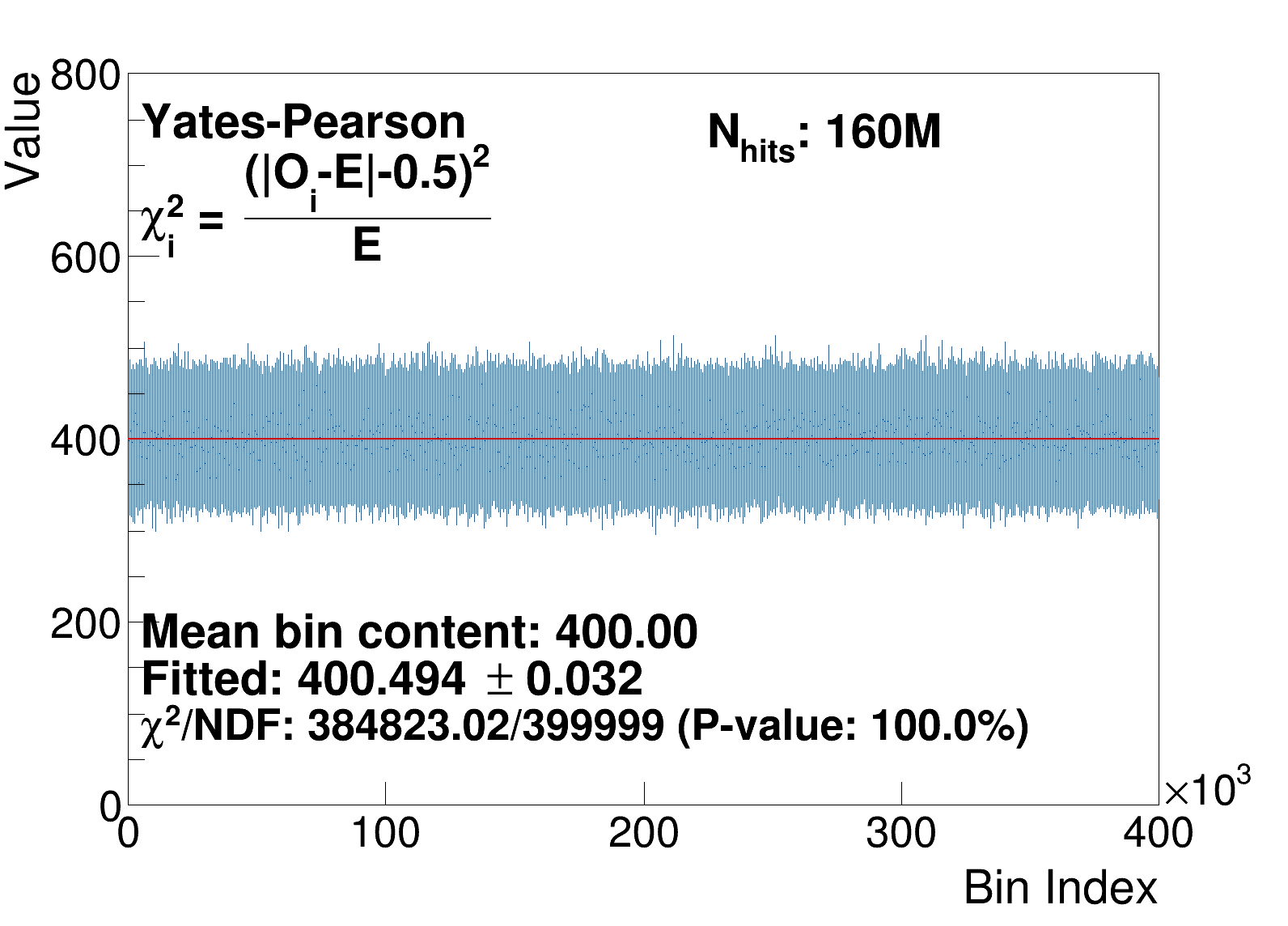}{Yates-Pearson, Eq.~\eqref{e:YatesMod}}\\[1ex]
    \fitpanel{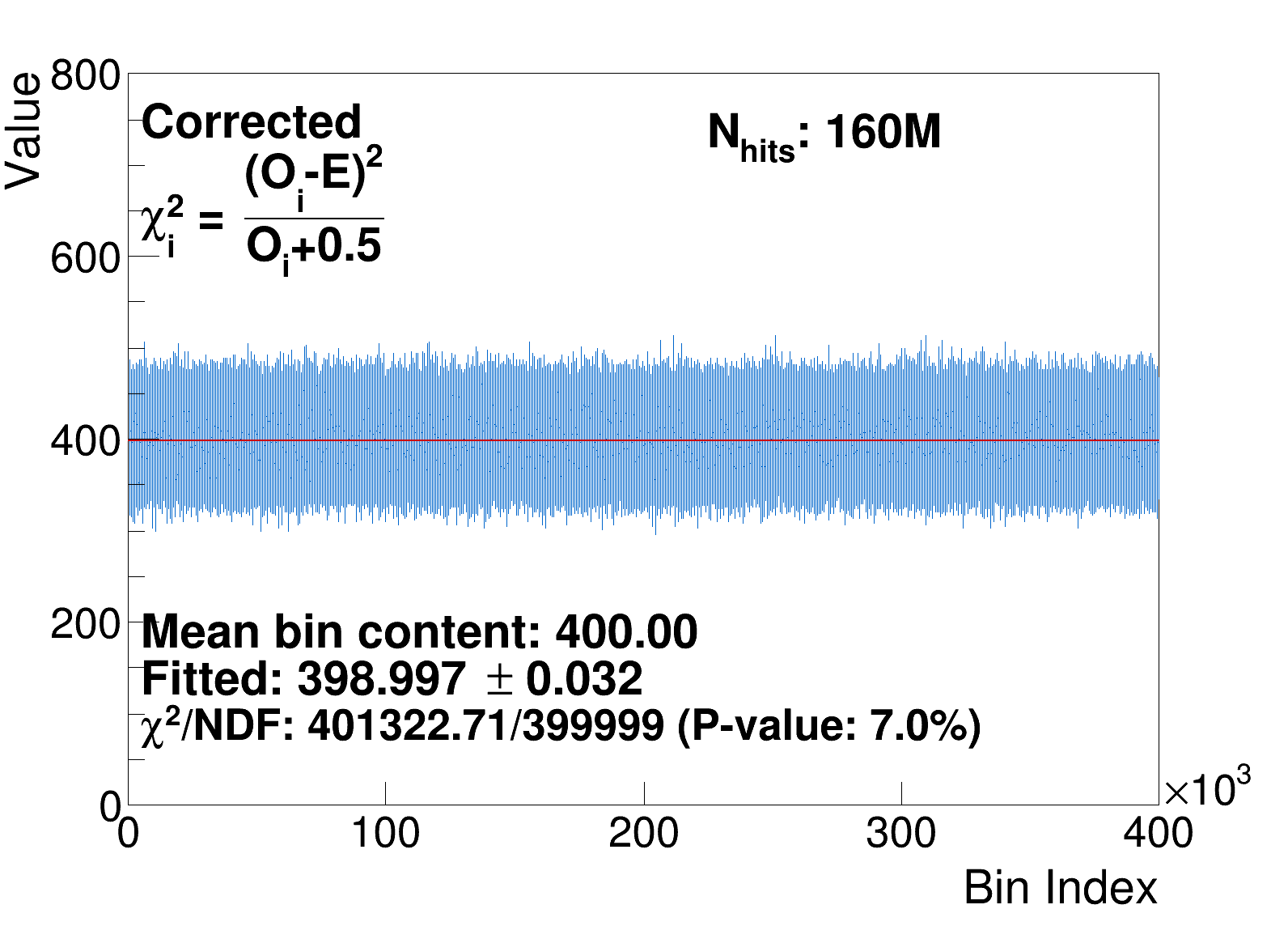}{Shifted variance, Eq.~\eqref{e:corr}}\hfill
    \fitpanel{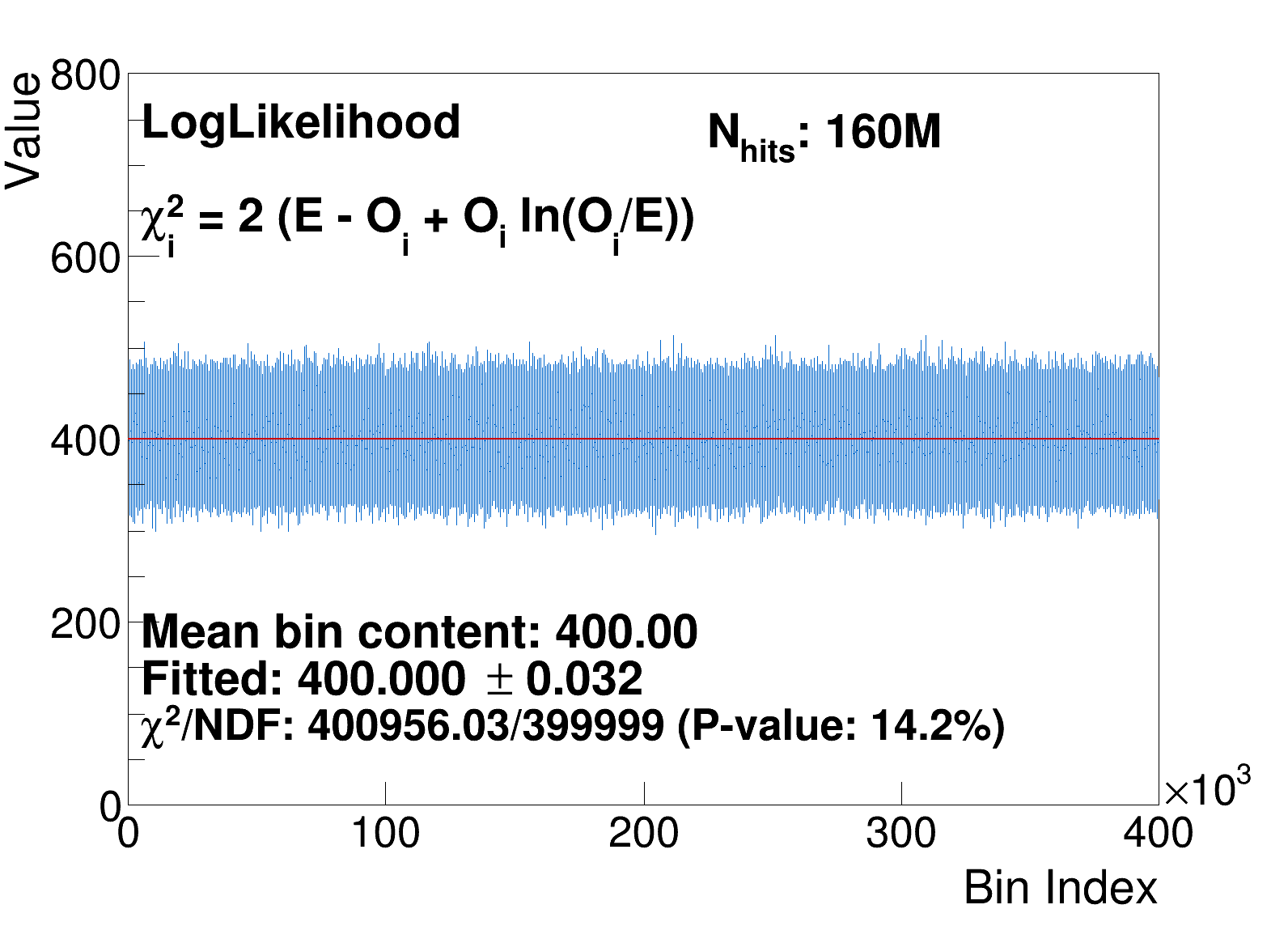}{Log-likelihood, Eq.~\eqref{e:LL}}
    \caption[]{(Continued.) $N_{\text{hits}} = 160$M ($\lambda = 400$).}
\end{figure}

\begin{figure}[p]
    \centering
    \fitpanel{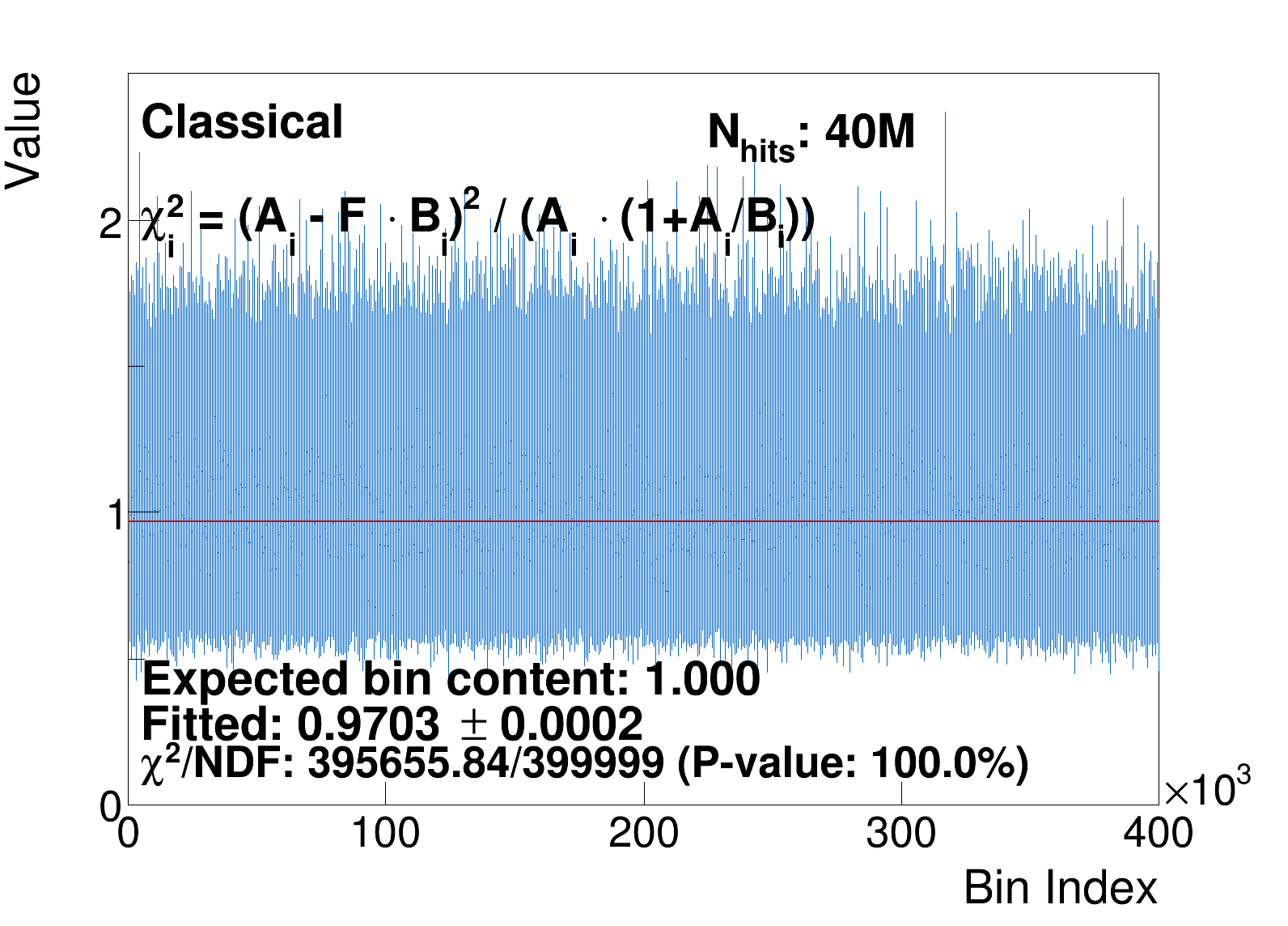}{Neyman, Eq.~\eqref{e:default_ratio}}\hfill
    \fitpanel{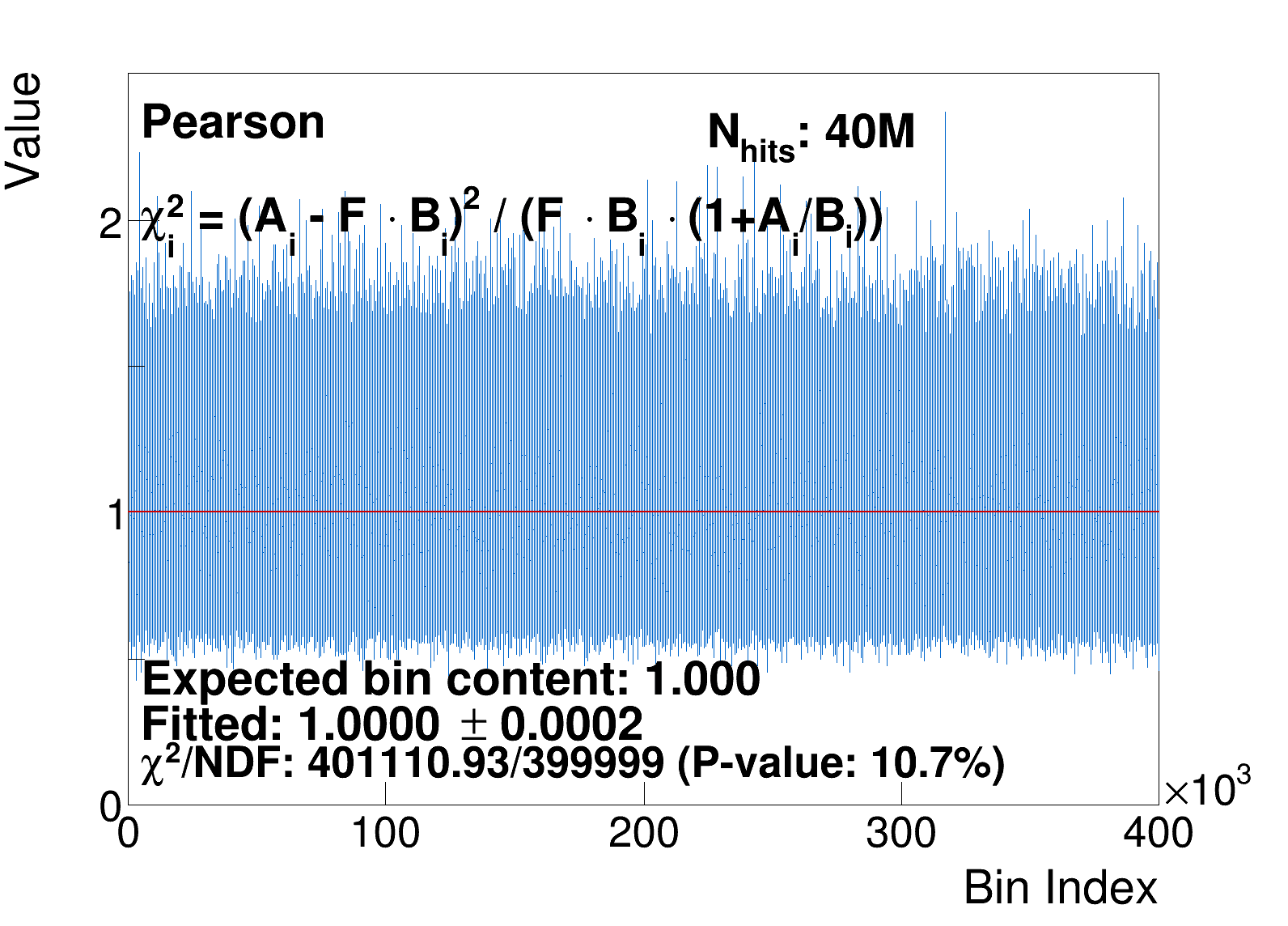}{Pearson, Eq.~\eqref{e:Pearson_ratio}}\\[1ex]
    \fitpanel{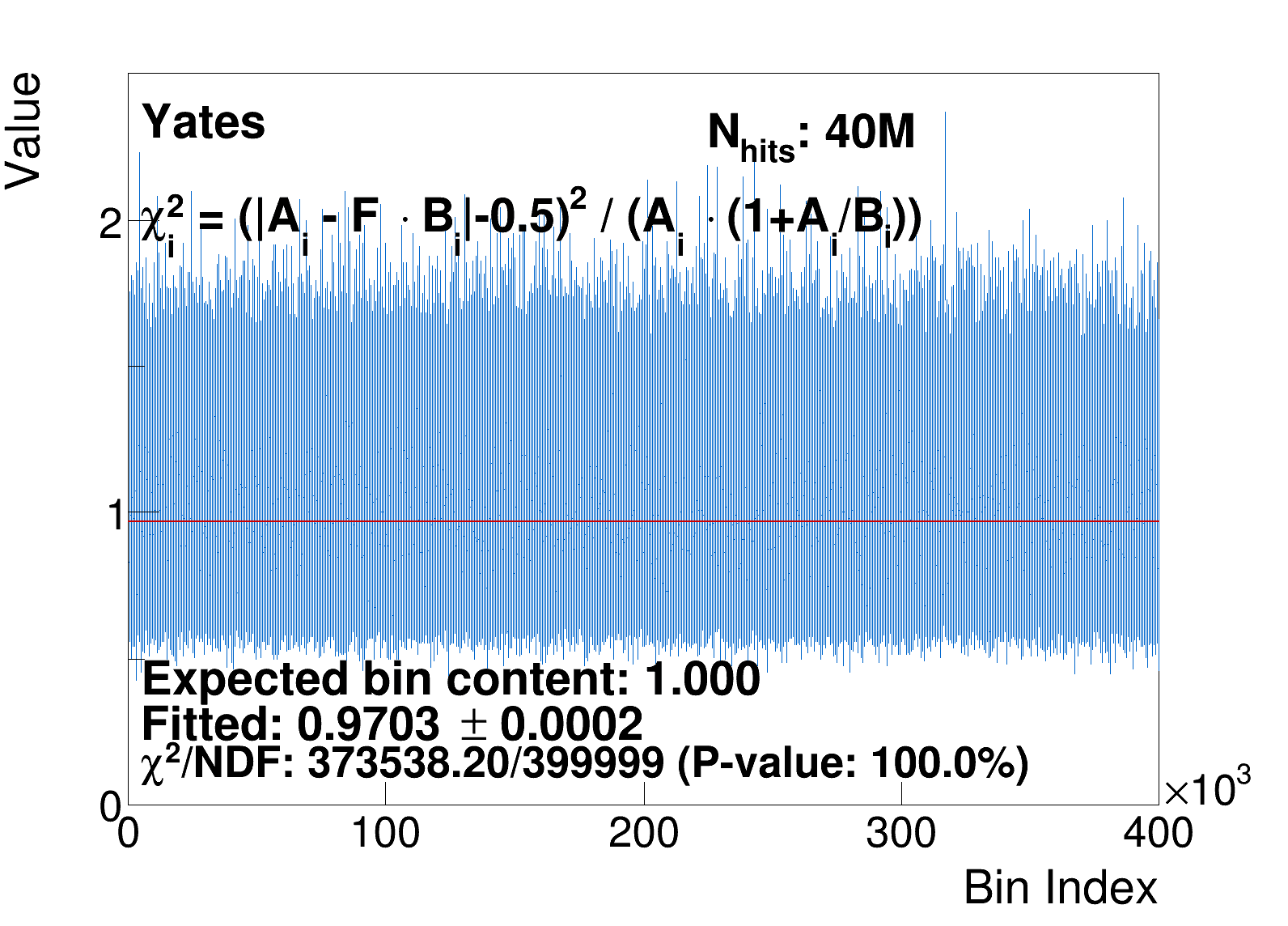}{Yates, Eq.~\eqref{e:Yates_ratio}}\hfill
    \fitpanel{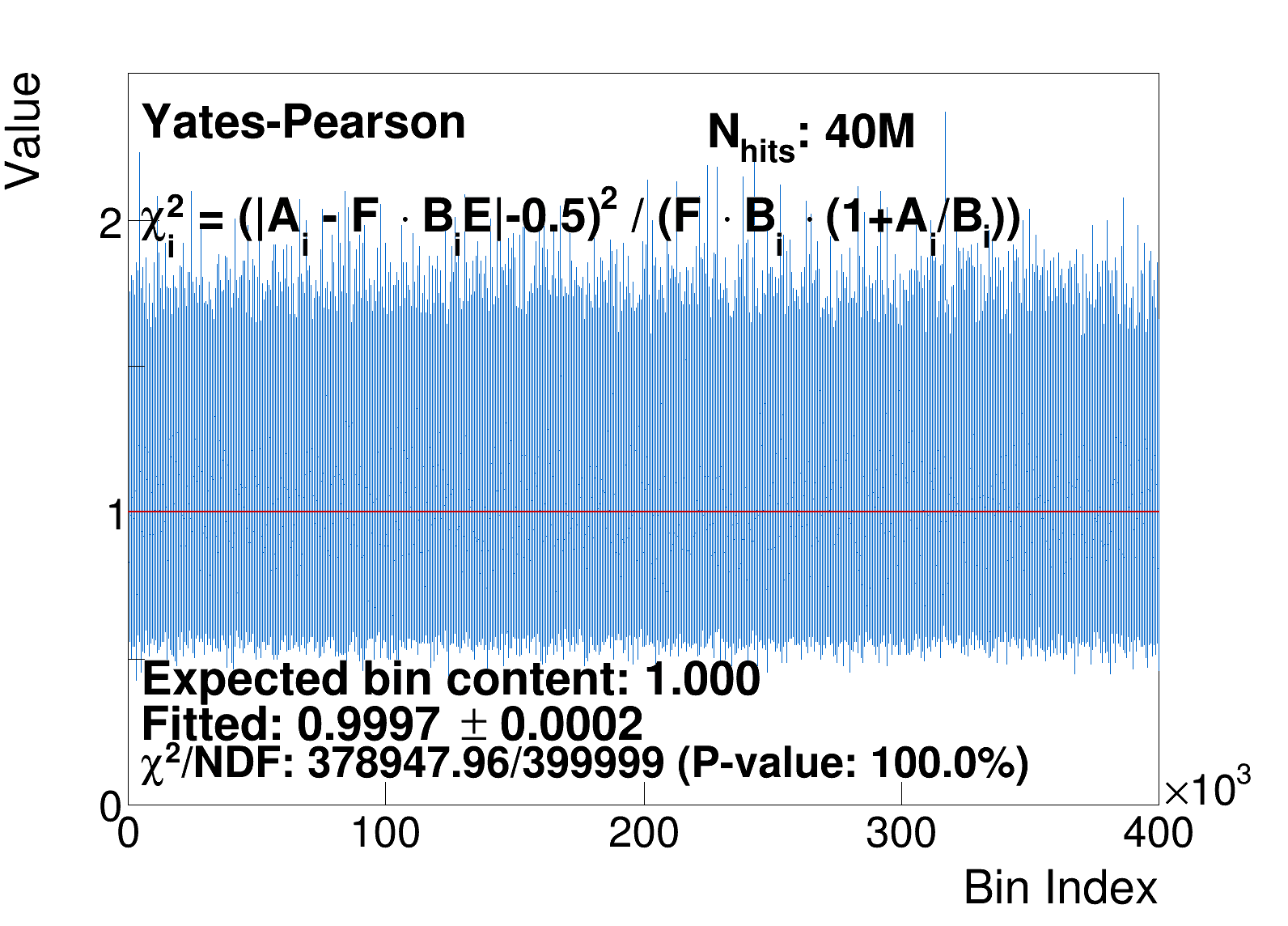}{Yates-Pearson, Eq.~\eqref{e:YatesMod_ratio}}\\[1ex]
    \fitpanel{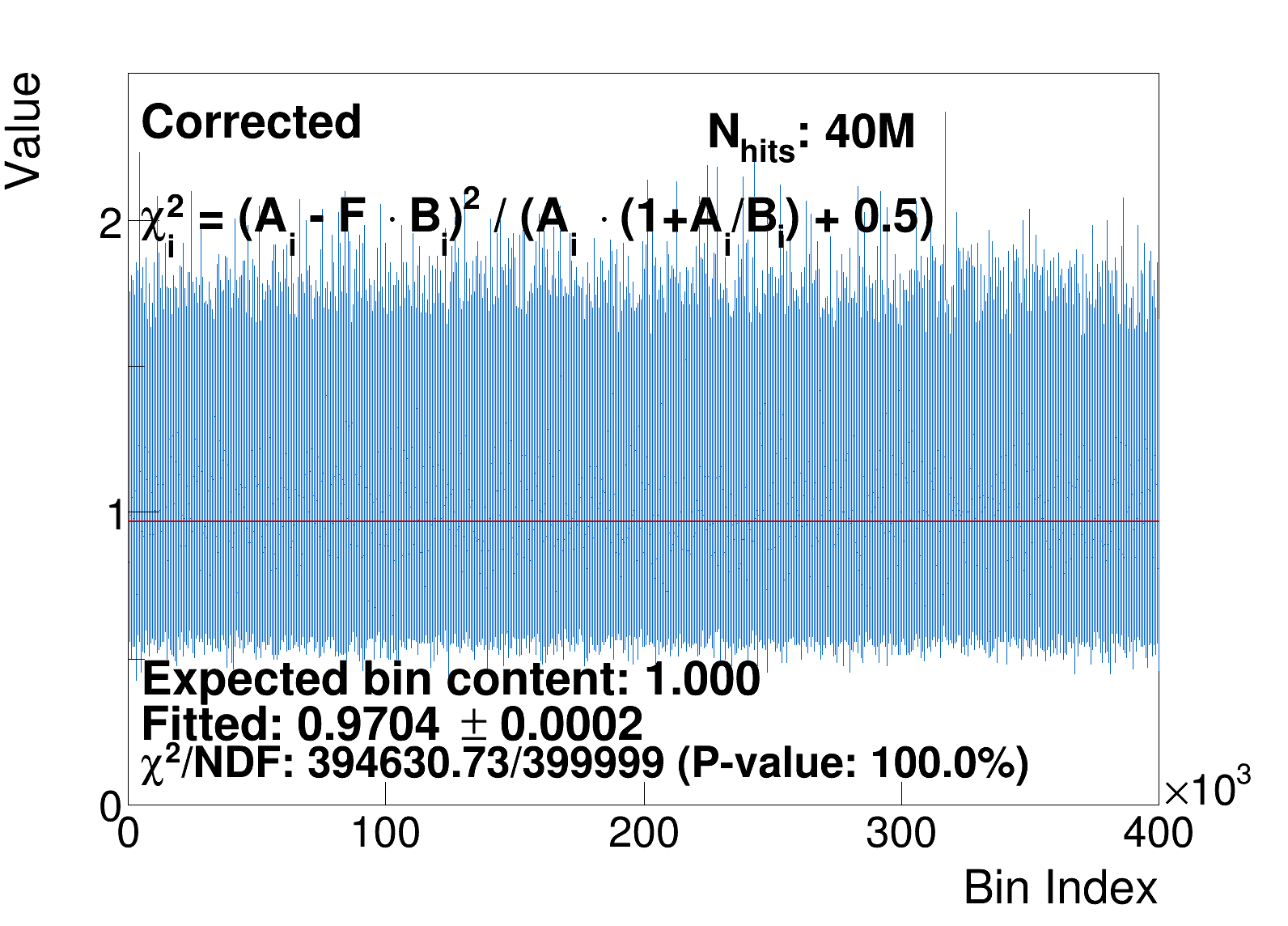}{Shifted variance, Eq.~\eqref{e:corr_ratio}}\hfill
    \fitpanel{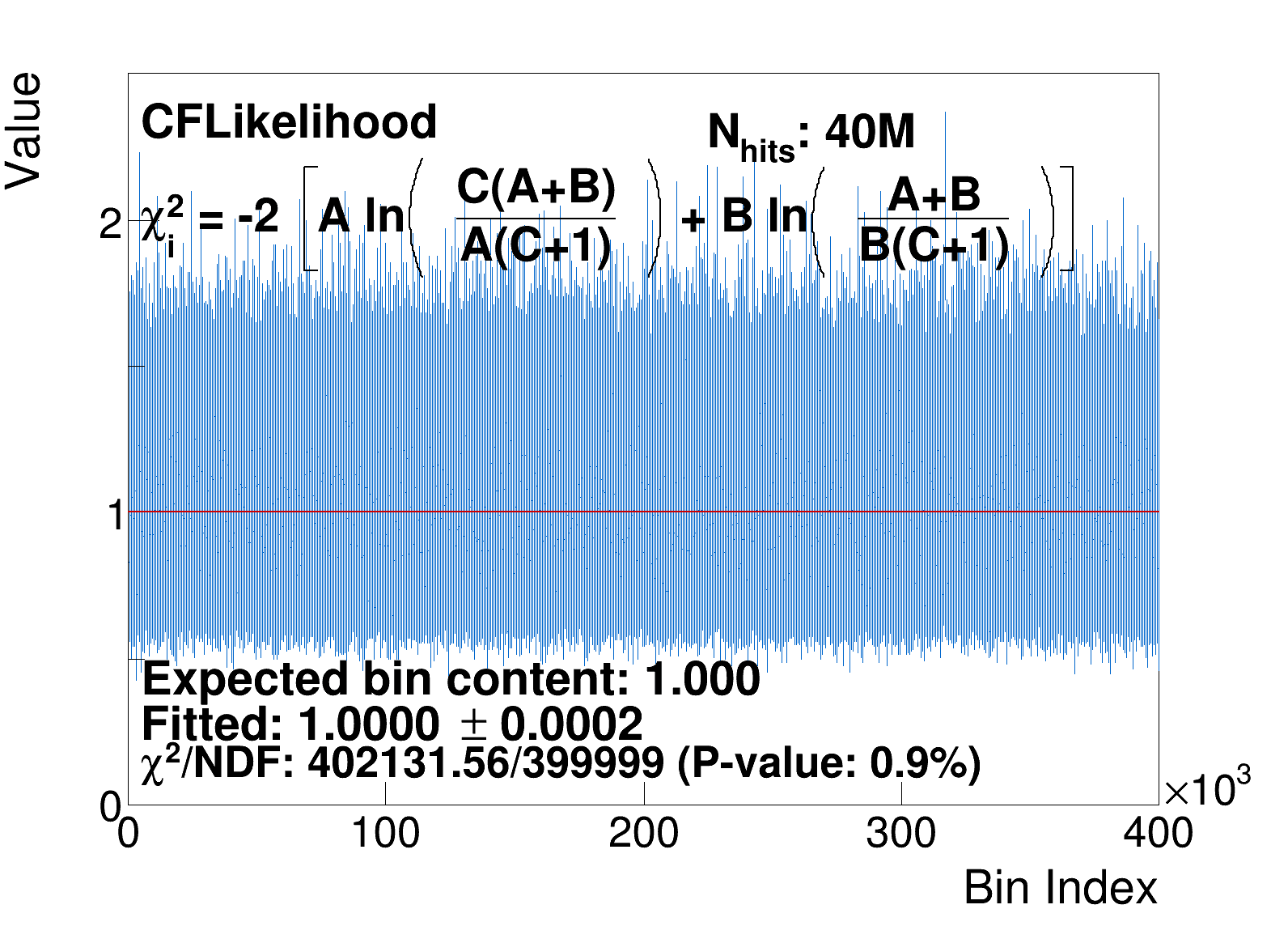}{CF likelihood, Eq.~\eqref{e:cf_ll}}
    \caption{Fits of a constant to the ratio of two statistically independent histograms with $N_{\text{bins}} = 400\,000$ bins each, filled with the same number of uniformly distributed random entries, using the estimators of Sec.~\ref{ss:neyman}--\ref{ss:cfl}. All panels on a page show the same pair of histograms. This page: $N_{\text{hits}} = 40$M ($\lambda = 100$).}
    \label{fig:ratio}
\end{figure}

\begin{figure}[p]
    \ContinuedFloat
    \centering
    \fitpanel{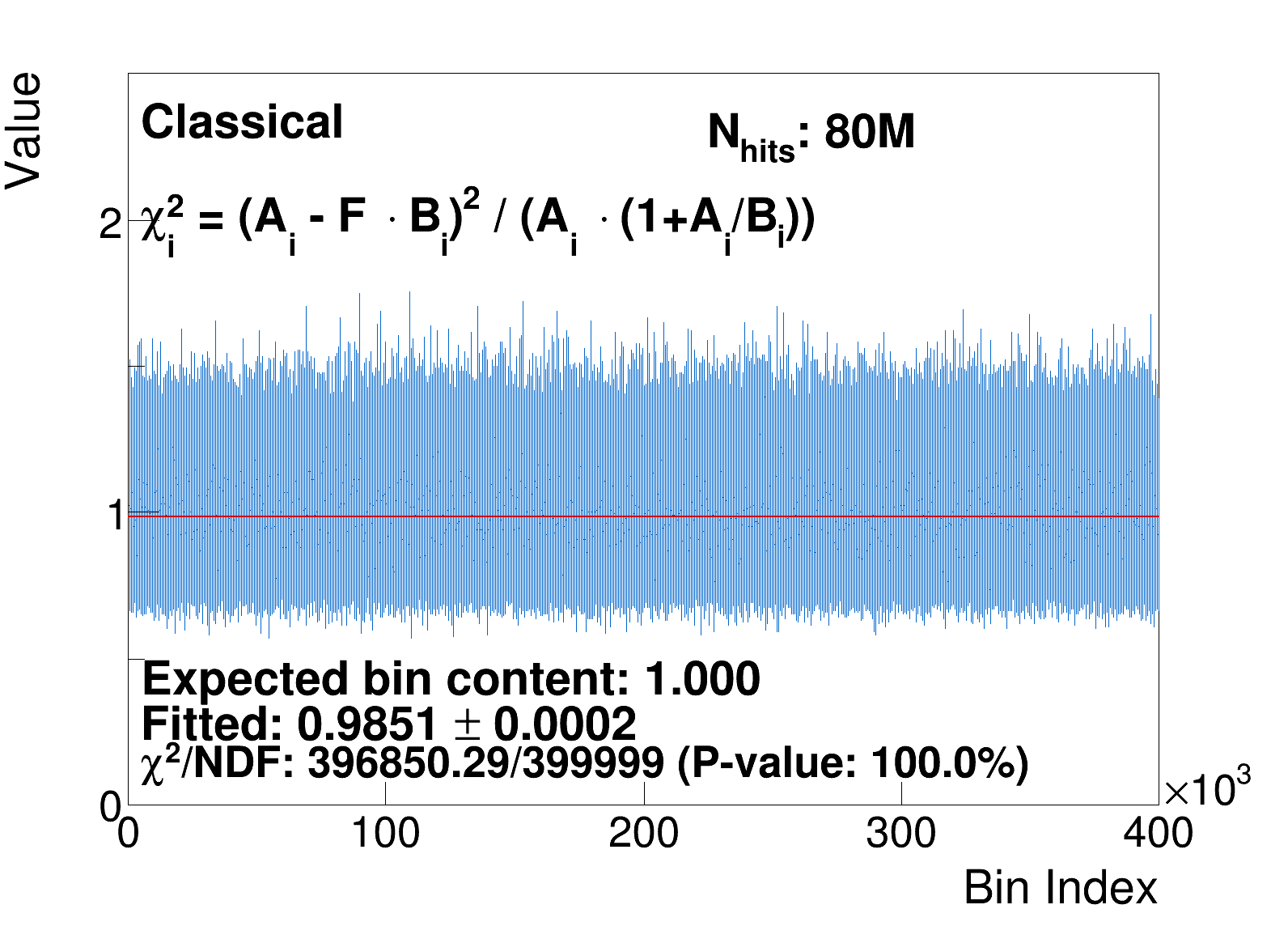}{Neyman, Eq.~\eqref{e:default_ratio}}\hfill
    \fitpanel{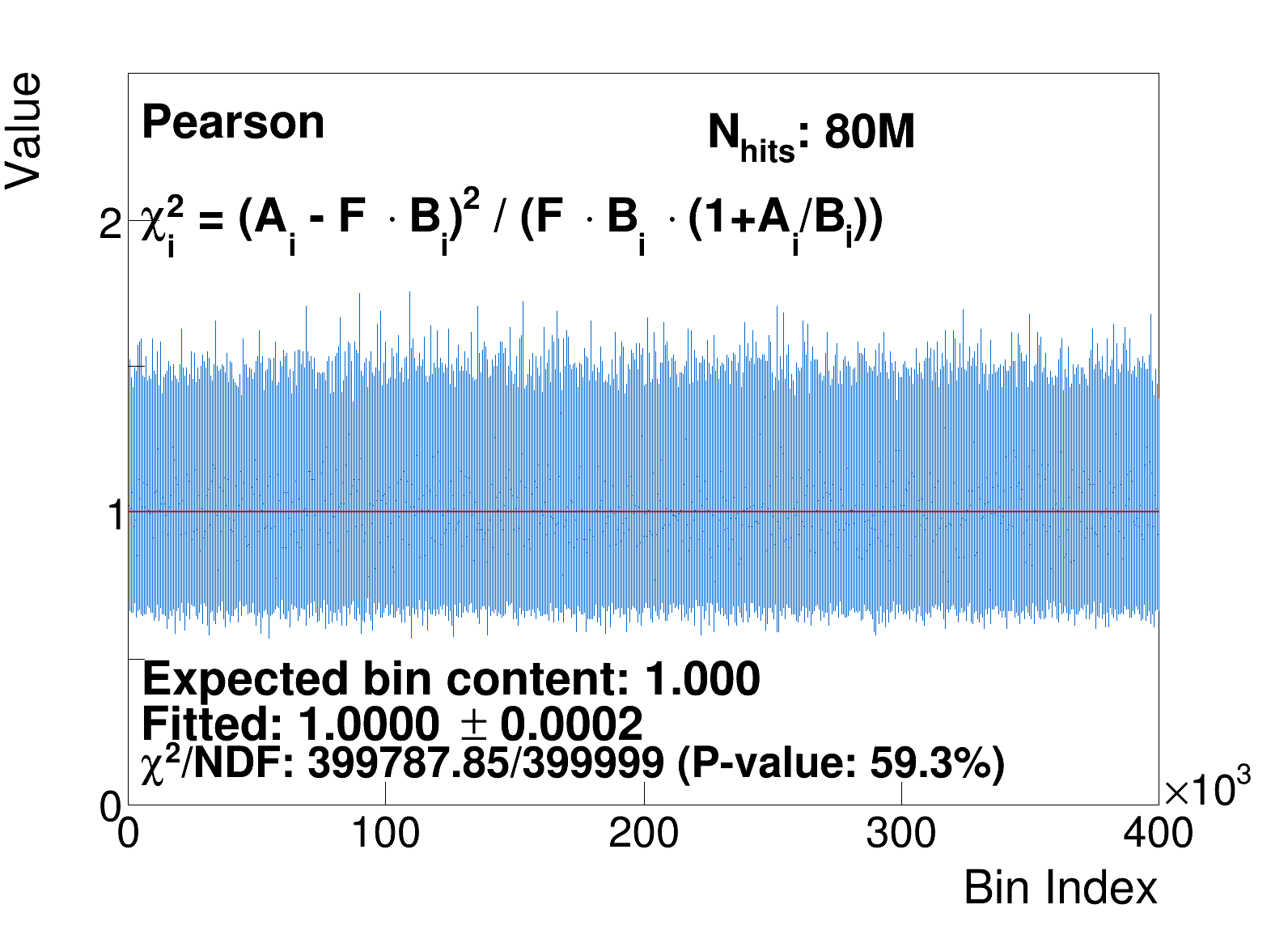}{Pearson, Eq.~\eqref{e:Pearson_ratio}}\\[1ex]
    \fitpanel{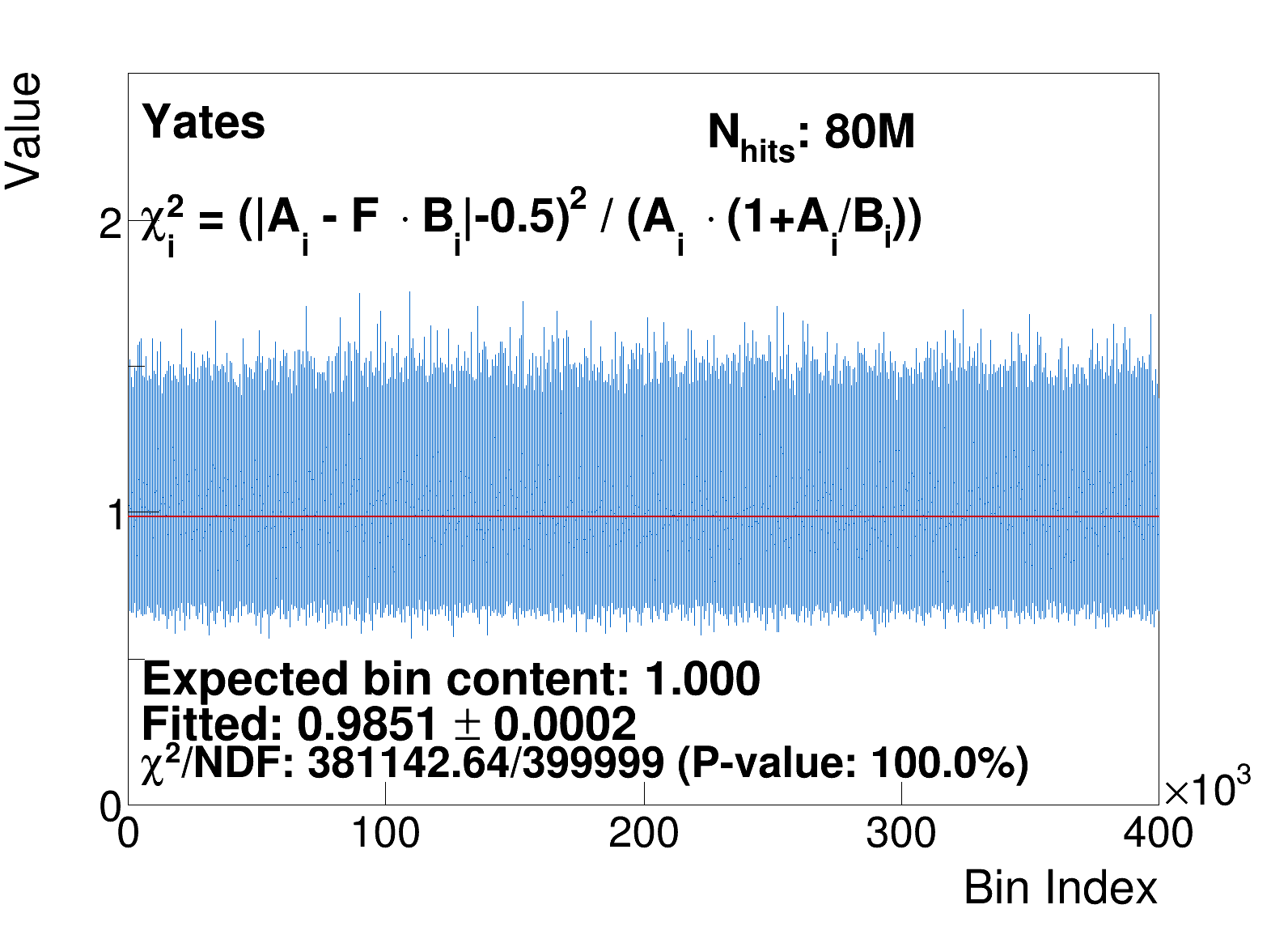}{Yates, Eq.~\eqref{e:Yates_ratio}}\hfill
    \fitpanel{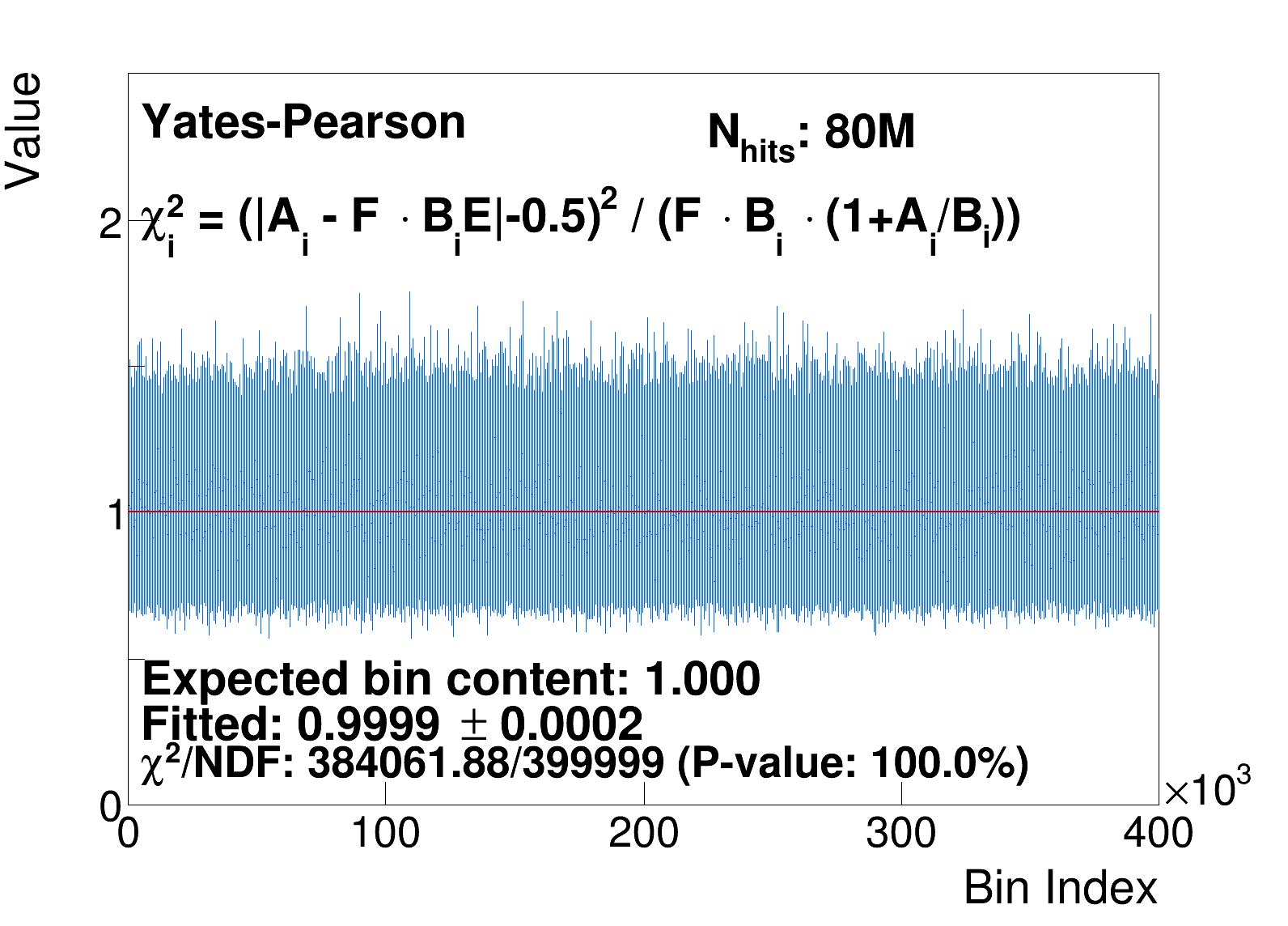}{Yates-Pearson, Eq.~\eqref{e:YatesMod_ratio}}\\[1ex]
    \fitpanel{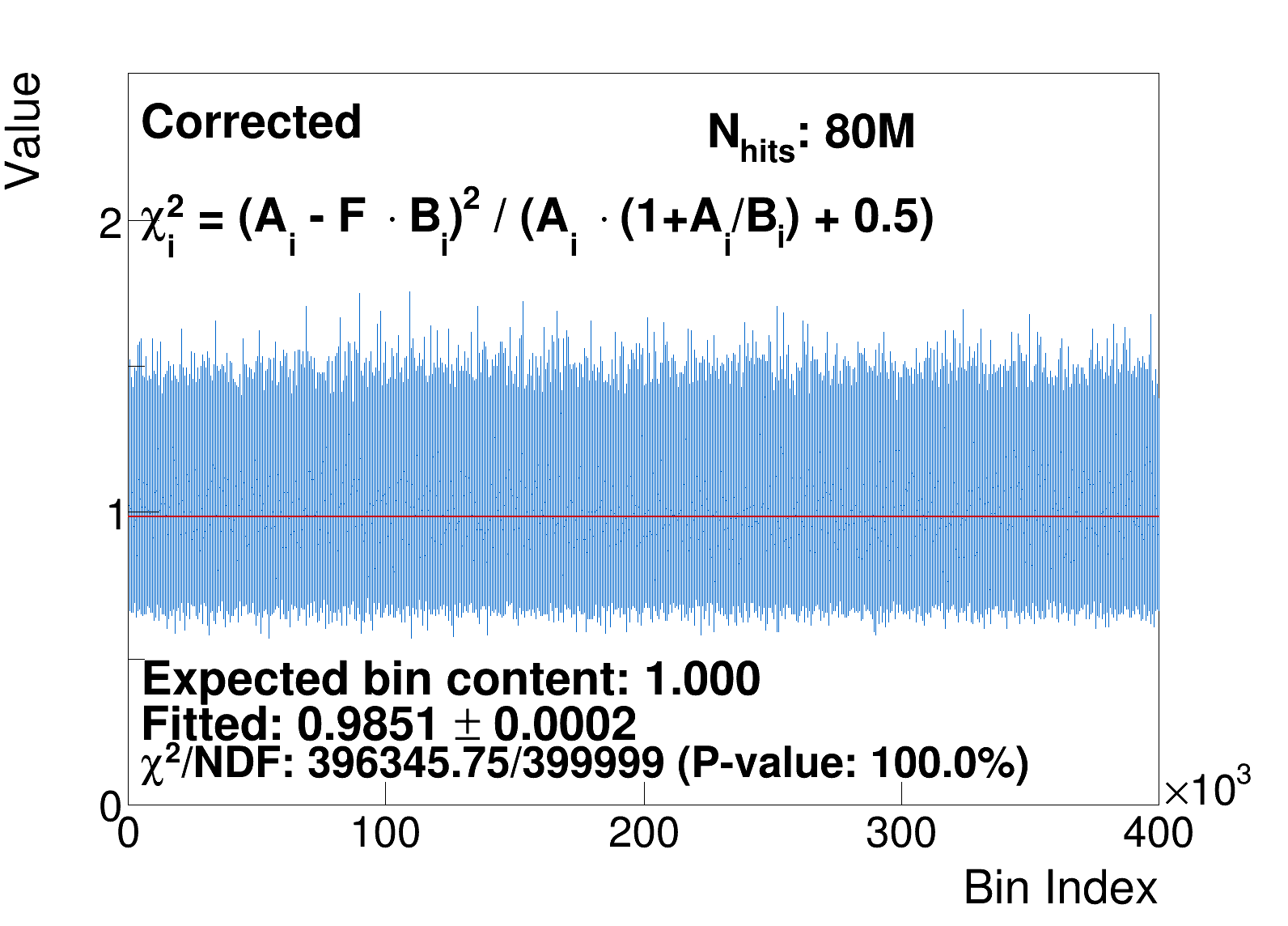}{Shifted variance, Eq.~\eqref{e:corr_ratio}}\hfill
    \fitpanel{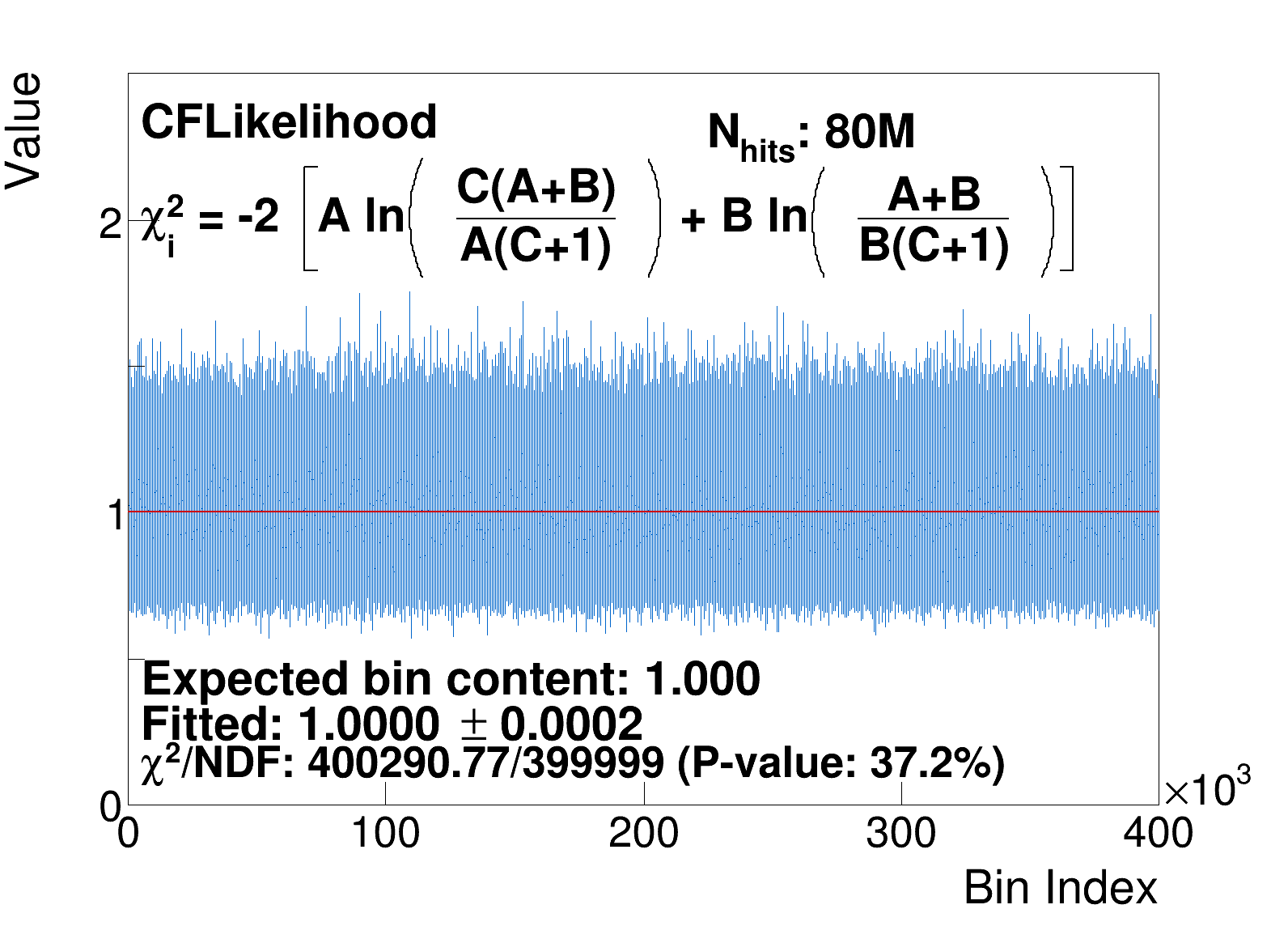}{CF likelihood, Eq.~\eqref{e:cf_ll}}
    \caption[]{(Continued.) $N_{\text{hits}} = 80$M ($\lambda = 200$).}
\end{figure}

\begin{figure}[p]
    \ContinuedFloat
    \centering
    \fitpanel{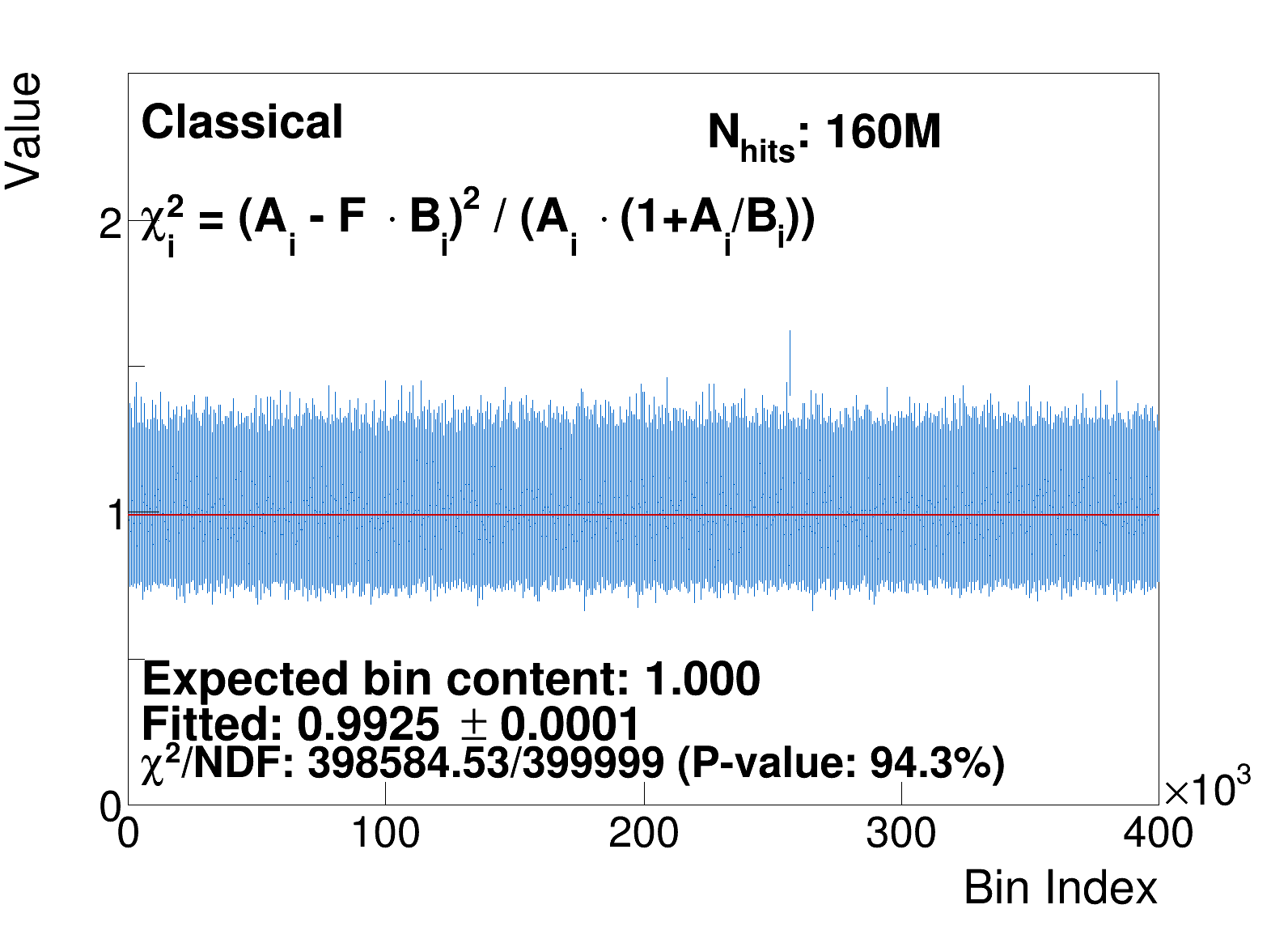}{Neyman, Eq.~\eqref{e:default_ratio}}\hfill
    \fitpanel{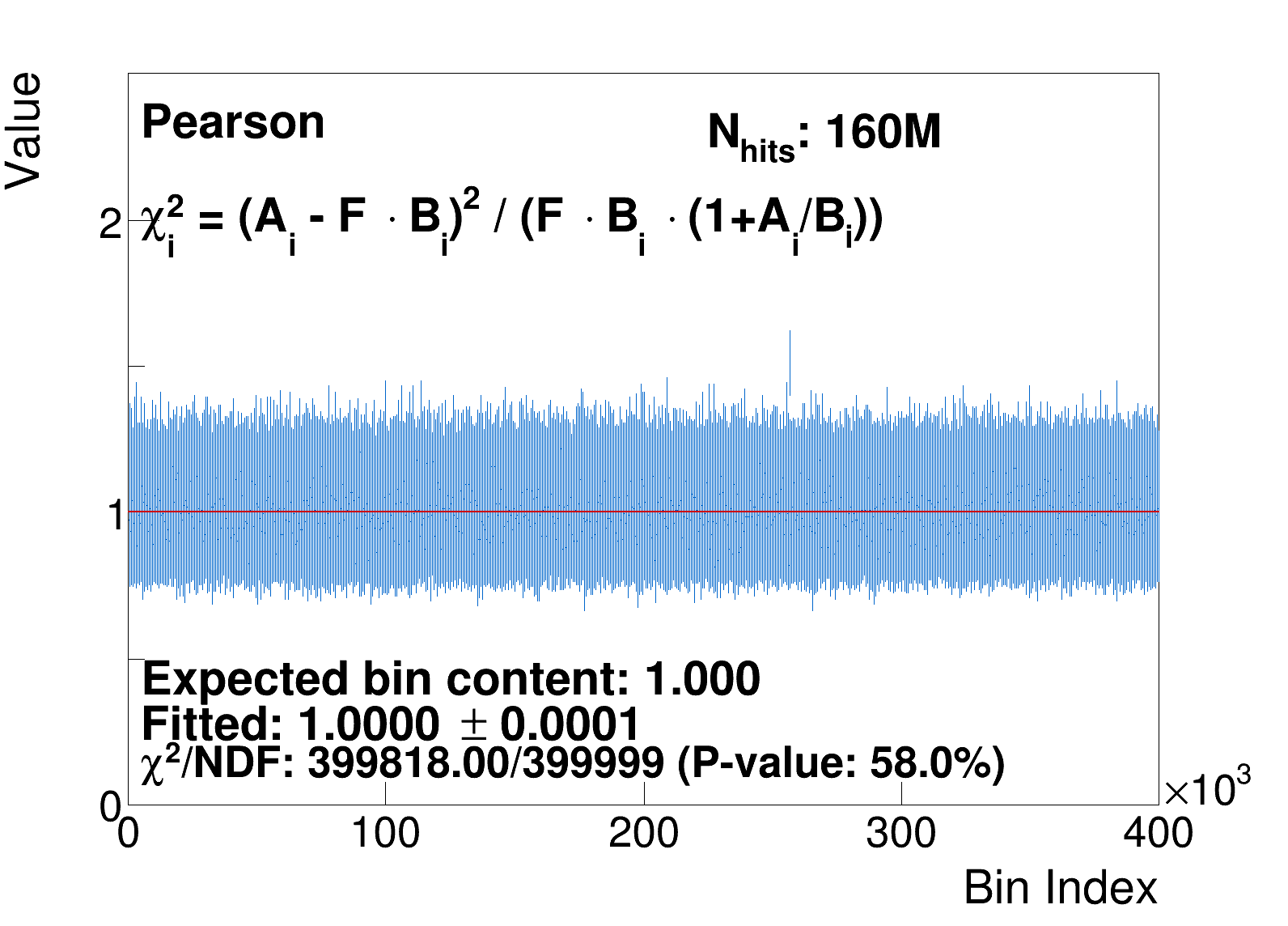}{Pearson, Eq.~\eqref{e:Pearson_ratio}}\\[1ex]
    \fitpanel{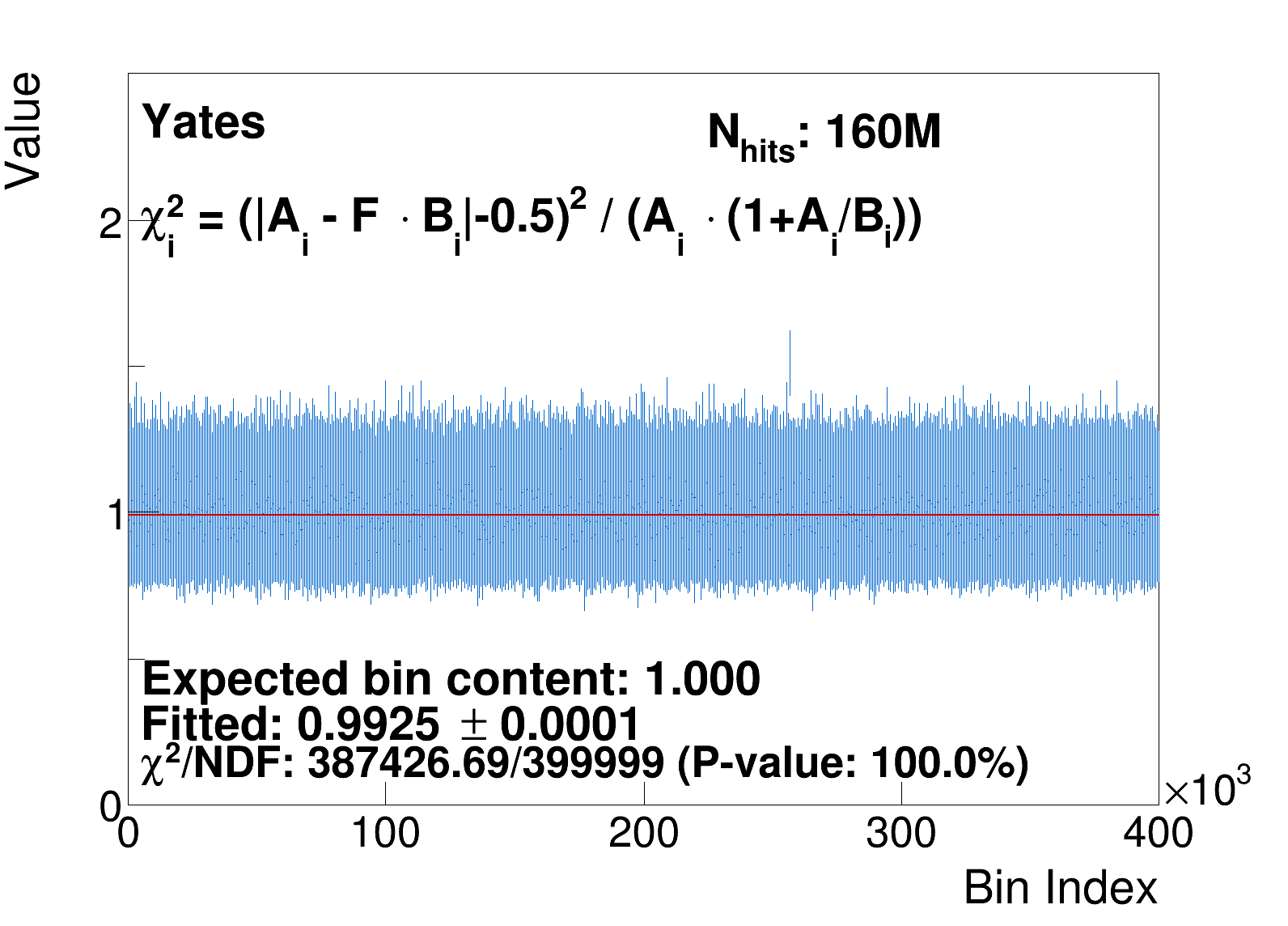}{Yates, Eq.~\eqref{e:Yates_ratio}}\hfill
    \fitpanel{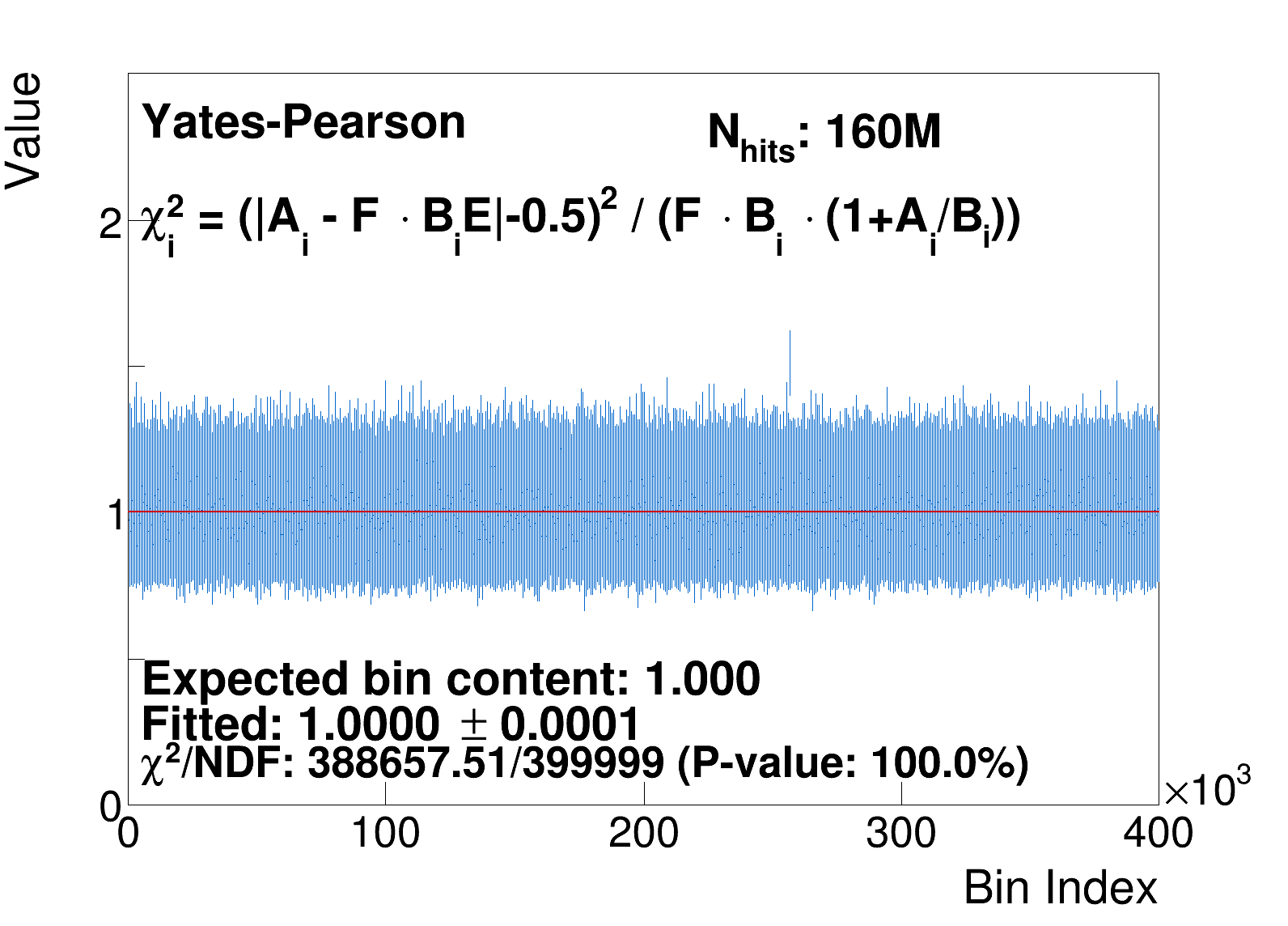}{Yates-Pearson, Eq.~\eqref{e:YatesMod_ratio}}\\[1ex]
    \fitpanel{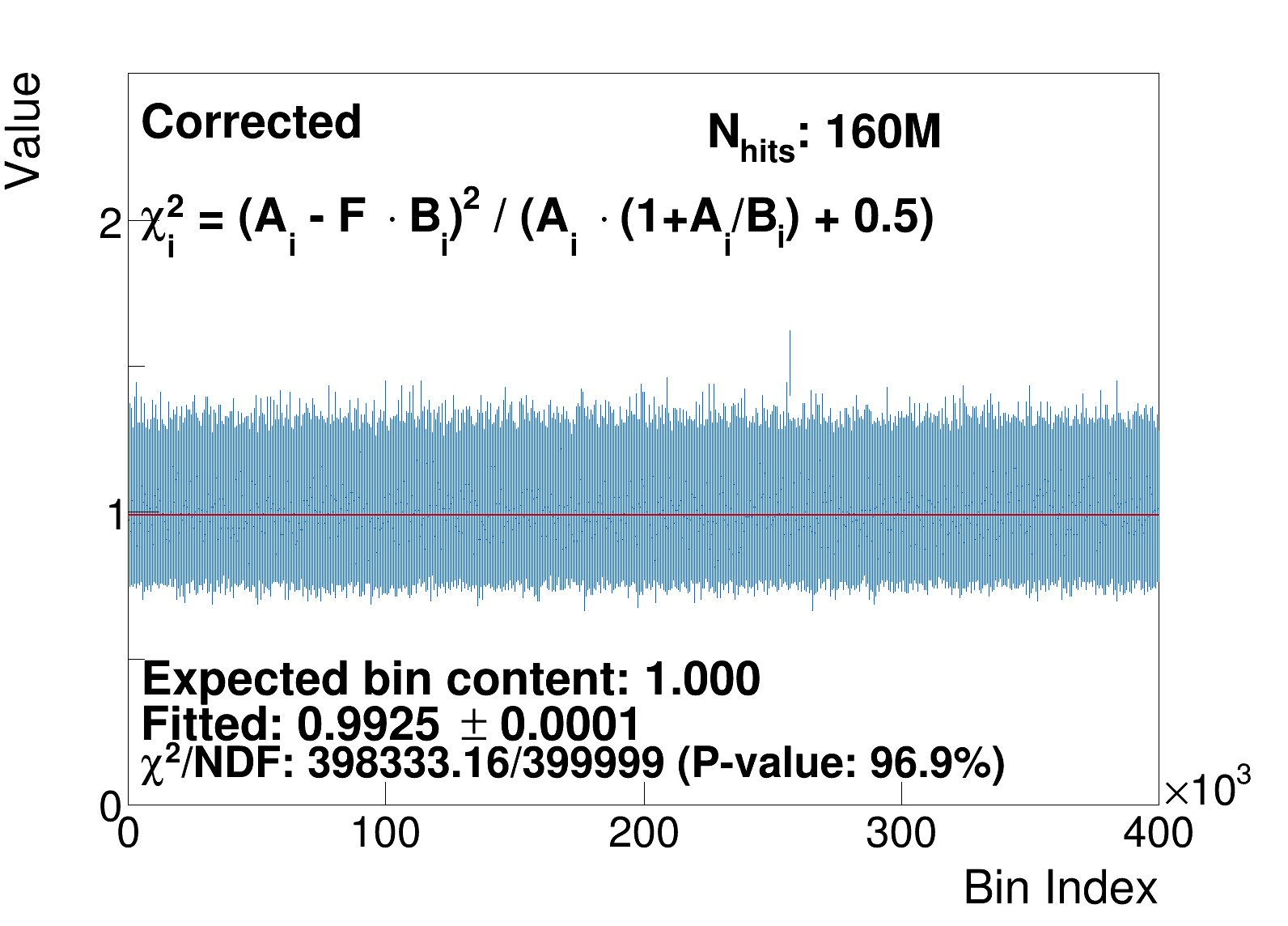}{Shifted variance, Eq.~\eqref{e:corr_ratio}}\hfill
    \fitpanel{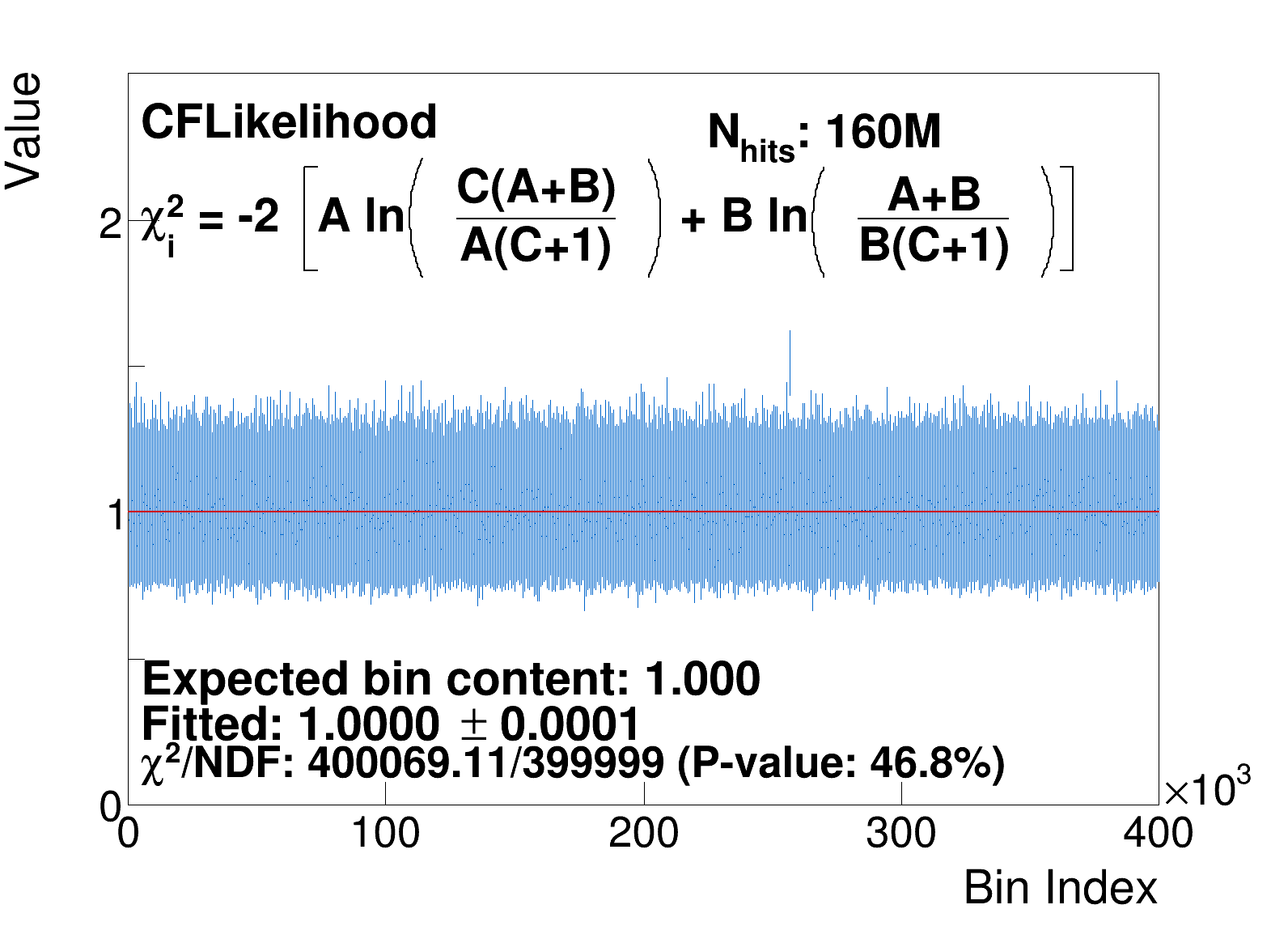}{CF likelihood, Eq.~\eqref{e:cf_ll}}
    \caption[]{(Continued.) $N_{\text{hits}} = 160$M ($\lambda = 400$).}
\end{figure}

\end{document}